\documentclass[preprint,groupedaddress,tightenlines,nofootinbib,floatfix,dvitex]{revtex4-1}

\usepackage{graphicx}
\usepackage{dcolumn}% Align table columns on decimal point
\usepackage{amssymb}
\usepackage{amsmath}
\usepackage{epstopdf}

\makeatletter
\newcommand{\figcaption}[1]{\def\@captype{figure}\caption{#1}}
\newcommand{\tblcaption}[1]{\def\@captype{table}\caption{#1}}
\makeatother

\def\simge{\mathrel{%
       \rlap{\raise 0.511ex \hbox{$>$}}{\lower 0.511ex \hbox{$\sim$}}}}
\def\simle{\mathrel{%
       \rlap{\raise 0.511ex \hbox{$<$}}{\lower 0.511ex \hbox{$\sim$}}}}

\begin{document}

\title{Canonical partition function and center symmetry breaking in finite density lattice gauge theories}

\author{Shinji Ejiri}
\affiliation{Department of Physics, Niigata University, Niigata 950-2181, Japan}

%\date{\today}
%\date{April 7, 2022}
\date{November 28, 2022}

\begin{abstract}
We study the nature of the phase transition of lattice gauge theories at high temperature and high density by focusing on the probability distribution function, which represents the probability that a certain density will be realized in a heat bath. 
The probability distribution function is obtained by creating a canonical partition function fixing the number of particles from the grand partition function. 
However, if the $Z_3$ center symmetry, which is important for understanding the finite temperature phase transition of $SU(3)$ lattice gauge theory, is maintained on a finite lattice, the probability distribution function is always zero, except when the number of particles is a multiple of three. 
For $U(1)$ lattice gauge theory, this problem is more serious. 
The probability distribution becomes zero when the particle number is nonzero.
This problem is essentially the same as the problem that the expectation value of the Polyakov loop is always zero when calculating with finite volume. 
In this study, we propose a solution to this problem.
We also propose a method to avoid the sign problem, which is an important problem at finite density, using the center symmetry.
In the case of $U(1)$ lattice gauge theory with heavy fermions, numerical simulations are actually performed, and we demonstrate that the probability distribution function at a finite density can be calculated by the method proposed in this study.
Furthermore, the application of this method to QCD is discussed.
\end{abstract}

\maketitle

%-------------------------------------------------------------------
\section{Introduction}
\label{intro}

Various results have been obtained so far in heavy-ion collision experiments to generate quark-gluon plasma.
In particular, recently, there has been a great deal of interest in the phase transition of QCD in the high-density region.
One of the most interesting topics is the verification that the first-order phase transition appears at high density, which is expected from the phenomenological discussion.
Since the critical point at which the low-density crossover changes to the first-order phase transition is the second-order phase transition, the fluctuation of the thermodynamic quantity of the heat bath increases as the particle density approaches the critical point.

In order to investigate the existence of such a critical point, the measurement of event-by-event fluctuations for conserved quantities such as net quark number and charge amount is attracting attention.
Furthermore, higher-order cumulants such as kurtosis and skewness have characteristic behaviors near the critical point, so they are also important physical quantities \cite{Ejiri:2005wq,Stephanov:2008qz,Asakawa:2009aj,Cheng:2008zh}.
On the other hand, in the numerical calculation of lattice QCD, which is the first-principles calculation, variance, skewness, and kurtosis can be calculated by the Taylor expansion method in the low-density region \cite{Allton:2003vx,Allton:2005gk,Ejiri:2009hq,DElia:2016jqh,Bazavov:2017dus,Borsanyi:2018grb,Bazavov:2020bjn,Bellwied:2021nrt}.
These are information about the shape of the probability distribution function (histogram) of the quark number or the charge.

The probability distribution function of the quark number can be obtained by computing the canonical partition function \cite{Hasenfratz:1991ax,Kratochvila:2005mk,deForcrand:2006ec,Ejiri:2008xt,Alexandru:2005ix,Li:2010qf,Li:2011ee,Danzer:2012vw,Fukuda:2015mva,Nakamura:2015jra,Bornyakov:2016wld,Wakayama:2018wkc}.
In this study, we calculate the probability distribution function of quark number.
The probability distribution function is a function that expresses the probability that a physical quantity will be a certain value.
Then, we aim to establish a method to determine the critical chemical potential (density) by numerical simulations of lattice QCD.
If there is a first-order phase transition, two states appear with equal probability at the phase transition point.
Therefore, if we calculate the probability distribution function and investigate its shape, we can determine whether the phase transition is first order, second order, or crossover.

The numerical calculation of the canonical partition function was performed in Ref.~\cite{Ejiri:2008xt} by a saddle point approximation in two flavor lattice QCD.
In the high-density region, the probability distribution function at the phase transition point has two peaks, indicating that the first-order phase-transition region exists.
However, since it is a finite density system, the sign problem arises in the calculation of the canonical partition function.
In Ref.~\cite{Ejiri:2008xt}, the sign problem is avoided by introducing an approximation that assumes that the probability distribution of the complex phase of the quark determinant, which causes the sign problem, is a Gaussian function. 
This approximation can be justified in the high-temperature phase, but in the low-temperature phase, the validity of the approximation is only that the actual numerical result is similar to the approximation function.

Moreover, the calculation of Ref.~\cite{Ejiri:2008xt} is implicitly limited to the case where the number of quarks is a multiple of three.
This is because, when calculated straightforwardly, the probability distribution function becomes zero except for multiples of three due to the $Z_3$-center symmetry.
The $Z_3$-center symmetry is an important symmetry for understanding the deconfinement phase transition of QCD as a spontaneous symmetry breaking.
If the symmetry is strictly maintained under the $Z_3$ center transformation, only states where the net quark number is zero or a multiple of three are realized.
It is consistent with the existence of only mesons (quark and antiquark) and baryons (three quarks) in the confined phase where symmetry is maintained.
The deconfinement phase is realized by the spontaneous symmetry breaking, which allows the quark to exist alone.
However, the symmetry is not broken in the actual numerical calculation with finite volume.
In other words, as long as there is symmetry, the number of quarks will always be a multiple of three.

In order to discuss phase transitions due to spontaneous symmetry breaking by a correct procedure, it is necessary to add an infinitesimal external field to break the symmetry and investigate the behavior in the limit of infinite volume.
We will discuss this issue in detail in Sec.~\ref{sec:center}. 
Moreover, in the case of $U(1)$ lattice gauge theory, the particle number probability distribution function is zero except when the net particle number is zero because of the $U(1)$ center symmetry.
This is reasonable in the confinement phase (disordered phase), but in the deconfinement phase (ordered phase), the net particle number can be nonzero.

This paper is organized as follows.
In the next section, we introduce the probability distribution function of the particle density, which is given by the canonical partition function.
Then, we discuss problems caused by the center symmetry in the calculation of the canonical partition function in Sec.~\ref{sec:center}. 
A method to calculate the canonical partition function by a saddle point approximation is explained in Sec.~\ref{sec:saddle}.
For the case of $U(1)$ lattice gauge theory, the sign problem in the canonical approach is avoided at the same time.
We demonstrate this method for the $U(1)$ gauge theory with heavy dynamical fermions. 
The results of the probability distribution function of the particle density are shown in Sec.~\ref{sec:u1gt}.
Then, the application to QCD is discussed in Sec.~\ref{sec:su3ap}.
Our conclusions are given in Sec.~\ref{sec:conclusion}.

\section{Canonical partition function and particle density probability distribution function}
\label{sec:canonical}

The relation between the grand partition function $Z_{GC}(T, \mu)$ with the chemical potential $\mu$ and the canonical partition function $Z_C(T,N)$ with the particle number $N$ is given by the fugacity expansion,
\begin{eqnarray}
Z_{GC}(T,\mu) = \sum_{N=-\infty}^{\infty} Z_C(T,N) e^{N \mu/T}
\label{eq:fex}
\end{eqnarray}
at each temperature $T$.
The left-hand side of this equation, $Z_{GC}(T, \mu)$, is the normalization factor of the Boltzmann weight and is classified by $N$ in the right-hand side.
Hence, $Z_C(T,N) e^{N \mu/T}$ can be regarded as a weight factor for each $N$, and the probability distribution of $N$ is in proportion to $Z_C(T,N) e^{N \mu/T}$.

Here, we introduce an effective potential as a function of the particle number $N$ by the canonical partition function $Z_C(T,N)$, 
\begin{eqnarray}
Z_{GC}(T,\mu) &=& \sum_{N=-\infty}^{\infty} e^{-V_{\rm eff}(T,N)}, \\
V_{\rm eff}(T,N) &=& -\ln Z_C(T,N) -\frac{N \mu}{T}.
\label{eq:veff}
\end{eqnarray}
We note that there is an uncertainty in adding a constant to $V_{\rm eff}$, which corresponds to an indefiniteness of a constant multiple of $Z_{GC}$.
The generation probability is maximized when the number of particles $N$ where $V_{\rm eff} (T,N)$ is the minimum.
We introduce the particle density $\rho$ as $\rho=N/V$, where $V$ is the spatial volume.
For sufficient large $V$, the minimum point is obtained from the derivative of $V_{\rm eff}(T, V \rho)$,
\begin{eqnarray}
\frac{1}{V} \frac{\partial V_{\rm eff} (T, V \rho)}{\partial \rho}  
= - \frac{1}{V} \frac{\partial \ln Z_C(T, V \rho)}{\partial \rho} - \frac{\mu}{T} = 0.
\end{eqnarray}
Hence, the derivative $- (1/V) \partial \ln Z_C / \partial \rho$ is the $\mu/T$ where the particle density with the maximum generation probability is $\rho$.

The grand partition function of $N_{\rm f}$ flavor QCD is defined by 
\begin{eqnarray}
Z_{GC}(T,\mu_1, \mu_2, \cdots, \mu_{N_{\rm f}}) = \int {\cal D} U \prod_{f=1}^{N_{\rm f}} \det M(\kappa_f, \mu_f/T) \ e^{-S_g}, 
\end{eqnarray}
where $\kappa_f$ is the hopping parameter and $\mu_f$ is the chemical potential for the $f$th flavor. 
The action we study consists of the gauge action and the fermion action,
\begin{eqnarray}
S = S_g + S_q,
\label{eq:Stot}
\end{eqnarray} 
where the gauge action is the standard plaquette action given by  
\begin{eqnarray}
S_g = -6 N_{\rm site} \,\beta \, \hat{P}
\label{eq:Sg}
\end{eqnarray}
with $N_{\rm site} = N_s^3 \times N_t$ the space-time lattice volume
and $\hat{P}$ the plaquette operator.
For the case of the standard plaquette gauge action of $SU(N_c)$ gauge theory, 
the gauge coupling parameter is $\beta = 2N_c/g^2$ and 
\begin{eqnarray}
\hat{P}= \frac{1}{6 N_c N_{\rm site}} \sum_{x,\,\mu < \nu} 
 {\rm Re \ tr} \left[ U_{x,\mu} U_{x+\hat{\mu},\nu}
U^{\dagger}_{x+\hat{\nu},\mu} U^{\dagger}_{x,\nu} \right] .
\label{eq:SgP}
\end{eqnarray} 
Here, $U_{x,\nu}$ is the link variable in the $\nu$ direction at site $x$ and 
$x+ \hat{\nu}$ the next site in the $\nu$ direction from $x$.
For fermions, we adopt the standard Wilson fermion action given by 
\begin{eqnarray}
S_q = \sum_{f=1}^{N_{\rm f}} \sum_{x,\,y} \bar{\psi}_x^{(f)} \, M_{xy} (\kappa_f, \mu_f/T) \, \psi_y^{(f)} ,
\label{eq:Sq}
\end{eqnarray} 
where $\psi^{(f)}$ is the fermion field and $M_{xy}$ is the Wilson fermion kernel
\begin{eqnarray}
M_{xy} (\kappa_f,\mu_f/T) &=& \delta_{xy} 
-\kappa_f \left\{ \sum_{i=1}^3 \left[ (1-\gamma_{i})\,U_{x,i}\,\delta_{y,x+\hat{i}} + (1+\gamma_{i})\,U_{y,i}^{\dagger}\,\delta_{y,x-\hat{i}} \right]  \right. 
\nonumber \\
       & & \left. + e^{\mu_fa}(1-\gamma_{4})\,U_{x,4}\,\delta_{y,x+\hat{4}}+e^{-\mu_fa}(1+\gamma_{4})\,U_{y,4}^{\dagger}\,\delta_{y,x-\hat{4}} \right\} .
\label{eq:Mq}
\end{eqnarray} 
The lattice spacing is $a$. 
Since $T=(N_t a)^{-1}$, the dimensionless ratio $\mu_f/T = \mu_f a N_t$.

To discuss the chemical potential dependence of the fermion determinant, we perform hopping parameter expansion in the vicinity of the heavy fermion limit, $\kappa =0$.
(See e.g. Sec.~5.1.3 of Ref.~\cite{Montvay94} or Sec.~11 of Ref.~\cite{Rothe92}.)
For each flavor, we have 
\begin{eqnarray}
\ln \left[ \frac{\det M(\kappa,\mu/T)}{\det M(0,0)} \right]
\;=\; \sum_{n=1}^{\infty} \frac{1}{n!} 
\left[ \frac{\partial^{n} \ln \det M}{\partial \kappa^{n}} 
\right]_{\kappa=0} \kappa^{n} 
\;=\; \sum_{n=1}^{\infty} \frac{B_n}{n!}  \, \kappa^{n} , 
\label{eq:tayexp}
\end{eqnarray}
with $\det M(0,0) =1$ and 
\begin{eqnarray}
B_n &\equiv& \left[ \frac{\partial^n \ln \det M}{\partial \kappa^n} \right]_{\kappa=0}
= (-1)^{n+1} (n-1)! \, {\rm tr} 
\left[ \left( M^{-1} \, \frac{\partial M}{\partial \kappa} \right)^n \right]_{\kappa=0}
\nonumber\\
&=&(-1)^{n+1} (n-1)! \, {\rm tr} 
\left[ \left( \frac{\partial M}{\partial \kappa} \right)^n \right], 
\label{eq:derkappa}
\end{eqnarray}
where $(\partial M/\partial \kappa)_{xy}$ is the hopping term following $\kappa_f$ in the right-hand side of Eq.~(\ref{eq:Mq}). 
Nonvanishing contributions to the trace of Eq.~(\ref{eq:derkappa}) appear only when the product of the hopping terms forms a connected closed loop in the space-time.
Therefore, $B_n$ are the sum of connected $n$-step Wilson loops \cite{Saito:2011fs}.
We classify them into Wilson-loop-type terms and Polyakov-loop-type terms.
The former are independent of the boundary conditions, and the latter are closed by the antiperiodic boundary condition for the time direction.
The leading-order (LO) contribution consists of the smallest Wilson-loop-type term, plaquette $\hat{P}$, defined by Eq.~(\ref{eq:SgP}), and the smallest Polyakov-loop-type term, Polyakov loop $\hat{\Omega}$, defined by 
\begin{eqnarray}
\hat\Omega = \frac{1}{N_c N_s^3} \sum_{\vec{x}} {\rm tr} \left[ 
U_{\vec{x},4} U_{\vec{x}+\hat{4},4} U_{\vec{x}+2 \cdot \hat{4},4} 
\cdots U_{\vec{x}+(N_t -1) \cdot \hat{4},4} \right] , 
\label{eq:ploop}
\end{eqnarray}
where $\sum_{\vec{x}}$ means a summation over one time slice. 

Because of the antiperiodic boundary condition and gamma matrices in the hopping terms, 
up to the LO contributions for $SU(N_c)$ gauge theory \cite{Saito:2013vja}, Eq.~(\ref{eq:tayexp}) reads 
\begin{eqnarray}
\ln \left[ \prod_{f=1}^{N_{\rm f}} \det M(\kappa_f, \mu_f/T) \right] 
&=& 96 N_c N_{\rm site} \sum_{f=1}^{N_{\rm f}} \kappa_f^4 \hat{P} 
\nonumber \\ 
&& \hspace{-25mm} 
+ 2^{N_t+1} N_c N_s^3 \left(
\sum_{f=1}^{N_{\rm f}} \kappa_f^{N_t} e^{\mu_f/T} \hat\Omega
+\sum_{f=1}^{N_{\rm f}} \kappa_f^{N_t} e^{\mu_f/T} \hat\Omega^* 
\right) + \cdots .
\label{eq:proddetM}
\end{eqnarray}
The first term that is proportional to $\hat{P}$ can be absorbed into the gauge action by a shift $\beta \rightarrow \beta^*$ with 
\begin{eqnarray}
\beta^* = \beta + 16 N_c \sum_{f=1}^{N_{\rm f}} \kappa_f^4 .
\label{eq:betastar}
\end{eqnarray}
The next-leading order (NLO) contributions in the Wilson-loop-type terms are the six-step Wilson loops and are typical operators in improved gauge actions. 
Thus, the contributions of these NLO operators can also be absorbed by a shift of improvement parameters of the gauge action. 
Because a shift in improvement parameters only affects the lattice discretization errors, the six-step Wilson loop terms will not affect characteristic features of the system in the continuum limit. 
Moreover, the coefficients of the NLO terms are very small compared to improvement parameters of typical improved actions \cite{Wakabayashi:2021eye}.
In contrast, the terms proportional to the Polyakov-loop-type terms act like external magnetic fields in spin models and thus may change the nature of the phase transition. 
The higher-order contribution is discussed in Refs.~\cite{Ejiri:2019csa,Kiyohara:2021smr,Wakabayashi:2021eye}.
We moreover assume that $N_t$ is an even number in this study.

As seen in Eq.~(\ref{eq:proddetM}), the Wilson-loop-type terms do not depend on the chemical potential. 
Since each term of $U_{x,4}$ in Eq.~(\ref{eq:Mq}) always appears as a combination of $e^{\mu_fa} U_{x,4}$ or $e^{-\mu_fa} U^{\dag}_{x,4}$, 
the chemical potential dependence of the expansion term containing $n_+$ $U_{x,4}$ and $n_-$ $U^{\dag}_{x,4}$ is $e^{\mu_f a(n_+-n_-)}$.
Because $n_+$ and $n_-$ satisfy $n_+-n_-=N_t N$ for a closed loop with a winding number $N$,
Wilson-loop-type terms do not depend on the chemical potential, and Polyakov-loop-type terms closed by the antiperiodic boundary condition with the winding number $N$ are in proportion to 
$e^{N N_t \mu_f a} = e^{N \mu_f /T}$.
Thus, the expansion terms can be classified by the winding number $N$, 
\begin{eqnarray}
\ln \left[ \det M(\kappa,\mu_f/T) \right]
= \sum_{N=-\infty}^{\infty} C_{N}  \, e^{N \mu_f/T} . 
\label{eq:tayexpmu}
\end{eqnarray}
$C_{N} e^{N \mu_f/T}$ is the sum of the terms whose winding number is $N$ in the hopping parameter expansion.
If we classify the expansion terms by the winding number, the canonical partition function is obtained.
For example, when $N_{\rm f}=2$ with the same $\mu$ and $\kappa$, 
$Z_{C}(T, N)$ can be calculated from the equation,
\begin{eqnarray}
Z_{GC}(T,\mu) &=& \int\mathcal{D}U(\mathrm{det}M(\kappa, \mu/T))^{2}e^{-S_g}
= \int {\cal D}U \exp \left[ \sum_{m=-\infty}^{\infty} C_{m} e^{m \mu/T}
+ \sum_{n=-\infty}^{\infty} C_{n} e^{n \mu/T} \right] e^{-S_g} 
\nonumber \\ 
&& \hspace{-10mm} = \int {\cal D}U \sum_{l=0}^{\infty} \frac{1}{l!} \left[ \sum_{m=-\infty}^{\infty} C_{m} e^{m \mu/T}
+ \sum_{n=-\infty}^{\infty} C_{n} e^{n \mu/T} \right]^l e^{-S_g} 
\nonumber \\ 
&& \hspace{-10mm} \equiv 
\sum_{N=-\infty}^{\infty} \left[ \int {\cal D}U \, {\det}_{N} M \, e^{-S_g} \right] e^{N \mu/T} 
= \sum_{N=-\infty}^{\infty} Z_C(T,N) \, e^{N \mu/T}
\label{eq:detNM}
\end{eqnarray}
for each term of $e^{N \mu/T}$, if we perform path integral over gauge configurations.
Here, $\det_{N}M$ means the expansion term of $[ \det M(\kappa, \mu/T) ]^{N_{\rm f}}$ with respect to $N$.
Therefore, the fugacity expansion is the winding number $N$ expansion.
Then, the chemical potential enlarges the contributions having positive winding number and suppress the contributions having negative winding number.

\section{Center symmetry and the canonical partition function}
\label{sec:center}

\subsection{$SU(3)$ gauge theory}

The quenched QCD, in which no dynamical fermions are included, has $Z_3$ center symmetry.
The centers of $SU(3)$ group are $\{ I, \omega I$, $\omega^2 I \}$, where $\omega = e^{2\pi i/3}$ and $I$ is the $3 \times 3$ unit matrix.
The center transformation is defined as
\begin{eqnarray}
U_{(\vec{x},t),4} \to \omega \ U_{(\vec{x},t),4}
\end{eqnarray}
for all $\vec{x}$ in one time slice $t$.
Under the center transformation, the system is symmetric, i.e. the gauge action and integral measure ${\cal D}U$ are invariant.
However, the expectation value of Polyakov loop $\hat{\Omega}$ changes as 
\begin{eqnarray}
\langle \hat{\Omega} \rangle \to \omega \langle \hat{\Omega} \rangle.
\end{eqnarray}
If this symmetry is maintained, any expectation values do not change under the center transformation.
Then, $\langle \hat{\Omega} \rangle = \omega \langle \hat{\Omega} \rangle$ must be satisfied. 
Thus, $\langle \hat{\Omega} \rangle$ is zero.
Therefore, the Polyakov loop is regarded as an order parameter of the spontaneous breaking of the center symmetry.

Similarly, under the $Z_3$ center transformation, the canonical partition function of the particle number $N$, 
$Z_C(T,N)$, changes as 
\begin{eqnarray}
Z_C(T,N) \to \omega^N Z_C(T,N) = e^{2\pi Ni/3} Z_C(T,N), 
\label{eq:centerzc}
\end{eqnarray}
since $Z_C(T,N)$ is given as the sum of the expectation values of Polyakov-loop-type operator with a winding number of $N$ as discussed in the previous section.
Because the grand partition function is given by the fugacity expansion, 
$Z_{GC}(T,\mu)$ changes as 
\begin{eqnarray}
\sum_{N=-\infty}^{\infty} Z_C(T,N) e^{N \mu/T} \to 
\sum_{N=-\infty}^{\infty} Z_C(T,N) e^{N \mu/T +2\pi Ni/3},
\end{eqnarray}
and is not invariant under the center transformation when including dynamical fermions.
At the same time, this implies 
\begin{eqnarray}
Z_{GC}(T, \mu) = Z_{GC}(T, \mu +2\pi iT/3),
\end{eqnarray}
since $S_g$ and the integral measure do not change under the center transformation, where we extend the real $\mu$ to a complex number.
This symmetry is called Roberge-Weiss symmetry \cite{Roberge:1986mm}.

Moreover, this indicates 
$Z_C (T,N)=0$ expect when $N$ is a multiple of 3.
Since $\omega^3 =1$ and $1+\omega+\omega^2=0$,
\begin{eqnarray}
Z_{GC}(T, \mu) 
&=& \frac{1}{3} \left(Z_{GC}(T, \mu) + Z_{GC}(T, \mu +2\pi iT/3) +Z_{GC}(T, \mu +4\pi iT/3) \right) \nonumber \\
&=& \sum_{n=-\infty}^{\infty} Z_C(T, 3n) e^{3n \mu/T},
\end{eqnarray}
where $n$ is an integer.
This means that the number of particles can only exist in multiples of three.
This is reasonable in the confinement phase where only baryons and mesons exist. 
However, in the deconfinement phase, there should be states of all particle numbers.
The order parameter of spontaneous symmetry breaking has a nonzero expectation value when one vacuum is selected from multiple vacua in the limit of infinite volume.
Thus, the center symmetry is never broken in an actual simulation with finite volume.

\subsection{$U(1)$ gauge theory}

This problem is more serious in $U(1)$ gauge theory.
For the case of $U(1)$ gauge theory, the centers are all elements of $U(1)$ group because of  an Abelian group.
The link variable is given as 
$U_{x, \nu} =e^{i \theta_{x, \nu}}$, 
where $\theta_{x, \nu}$ is a real variable that takes the value of  
$-\pi < \theta_{x, \nu} \leq \pi$ on a link $(x, \nu)$. 
The gauge action is given by  
$S_g = -6 N_{\rm site} \,\beta \, \hat{P},$
with $\beta = 1/g^2$ the gauge coupling parameter
and $\hat{P}$ the plaquette operator, 
\begin{eqnarray}
\hat{P}= \frac{1}{6 N_{\rm site}} \sum_{x,\,\mu < \nu} 
\cos \left[ \theta_{x,\mu} + \theta_{x+\hat{\mu},\nu} 
- \theta_{x+\hat{\nu},\mu} - \theta_{x,\nu} \right] 
\label{eq:U1SgP}
\end{eqnarray} 
for the case of the standard plaquette action. 
When we perform the hopping parameter expansion, the LO contribution of the fermion determinant is given by
\begin{eqnarray}
\ln \left[ \prod_{f=1}^{N_{\rm f}} \det M(\kappa_f, \mu_f/T) \right] 
&=& 96 N_{\rm site} \sum_{f=1}^{N_{\rm f}} \kappa_f^4 \hat{P} 
\nonumber \\ 
&& \hspace{-25mm} 
+ 2^{N_t+1}N_s^3 \left(
\sum_{f=1}^{N_{\rm f}} \kappa_f^{N_t} e^{\mu_f/T} \hat\Omega
+i \sum_{f=1}^{N_{\rm f}} \kappa_f^{N_t} e^{\mu_f/T} \hat\Omega^* 
\right) + \cdots .
\label{eq:proddetMu1}
\end{eqnarray}
The Polyakov loop for $U(1)$ gauge theory is 
\begin{eqnarray}
\hat{\Omega}= \frac{1}{N_s^3} \sum_{\vec{x}} 
\exp \left[ i( \theta_{\vec{x},4} + \theta_{\vec{x}+\hat{4},4} 
+ \theta_{\vec{x}+2\hat{4},4} + \cdots + \theta_{\vec{x}+N_t \hat{4},4}) \right] .
\end{eqnarray} 

Under the $U(1)$ center transformation: 
$U_{(\vec{x},t),4} \to e^{i \eta} U_{(\vec{x},t),4}$ on one time slice $t$,
$Z_C(T,N)$ changes as 
\begin{eqnarray}
Z_C (T,N) \to e^{iN \eta} Z_C(T,N).
\end{eqnarray}
Since  $S_g$ and the integral measure are invariant, $Z_C (T,N) = e^{iN \eta} Z_C(T,N)$. 
Thus, the canonical partition function is 
\begin{eqnarray}
Z_C (T,N) = \frac{1}{2 \pi} \int_0^{2 \pi} e^{iN \eta} Z_C(T,N) \, d \eta
= 0, 
\end{eqnarray}
except for $N=0$.
This means that the existence probability of particles that interact with the gauge field is zero.
Moreover, the grand partition function is 
\begin{eqnarray}
Z_{GC}(T,\mu_q) = Z_C(T,0) 
\end{eqnarray}
and does not depend on the chemical potential due to the $U(1)$ center symmetry.
Since the center symmetry cannot be broken in numerical simulations performed with finite volume, the canonical partition function cannot be calculated correctly.

\subsection{Polyakov loop in $U(1)$ gauge theory}
\label{sec:centerC}

\begin{figure}[tb]
\begin{center}
\vspace{0mm}
\includegraphics[width=7.5cm]{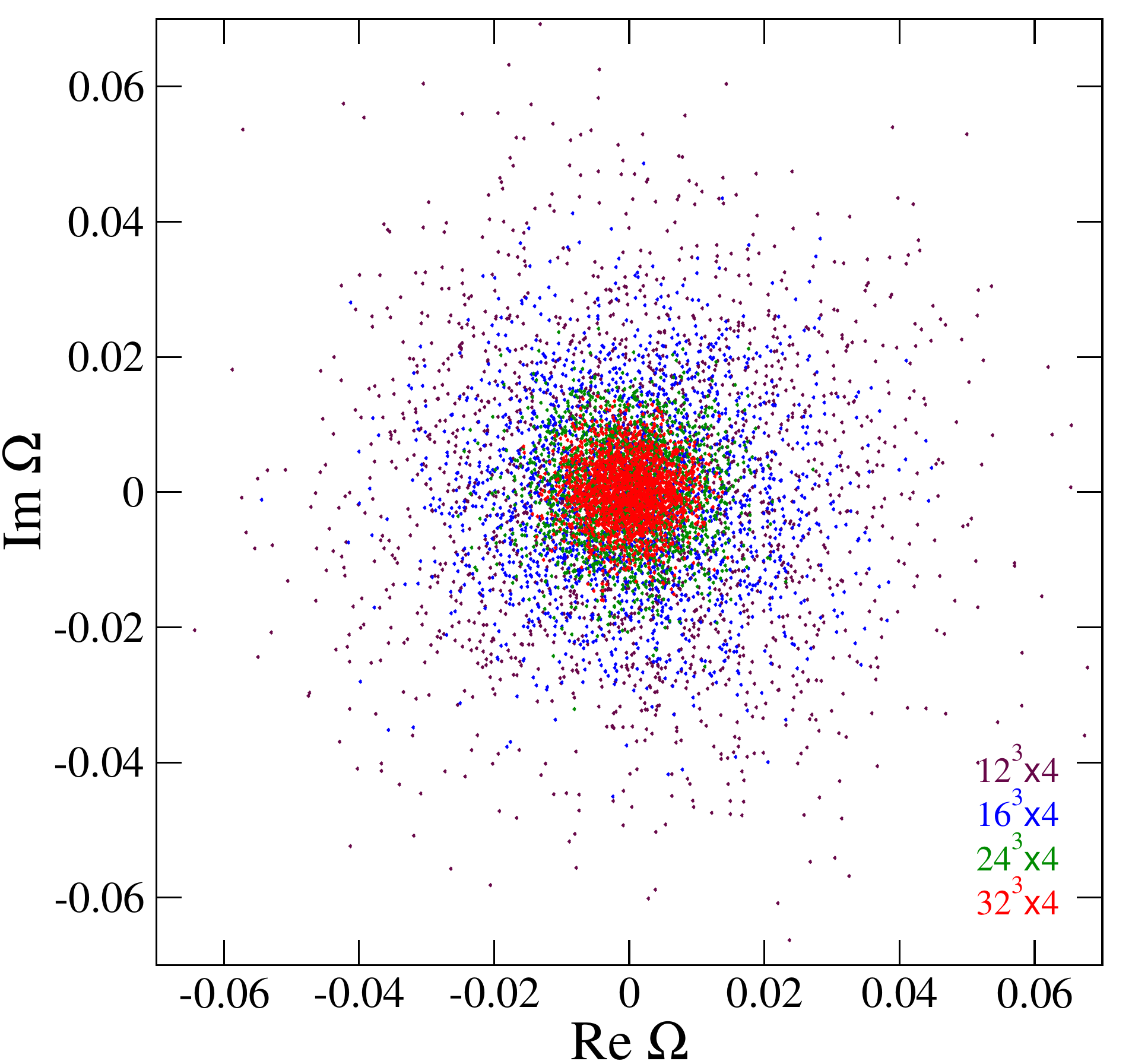}
\hspace{3mm}
\includegraphics[width=7.5cm]{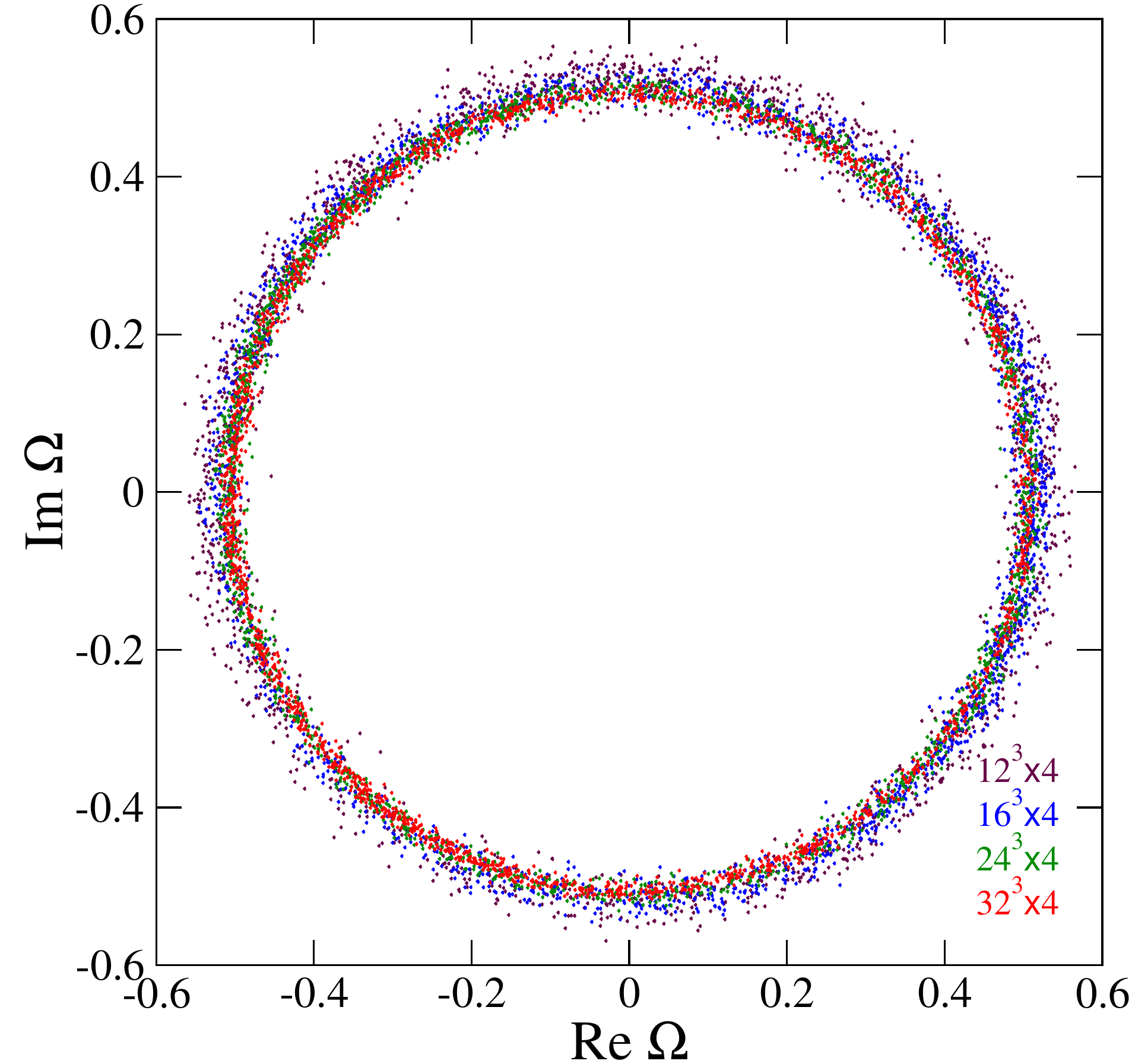}
\vspace{0mm}
\end{center}
\caption{Distribution of Polyakov loops in the complex plane generated by simulations of $U (1)$ lattice gauge theory at $\beta=0.90$ (left) and 1.10 (right). The temporal lattice size $N_t$ is 4.
Purple, blue, green and red symbols are the results of the spatial lattice size $N_s=12$, 16, 24, and 32, respectively.}
\label{fig:disU1}
\end{figure}

This problem is essentially the same as the problem that the expectation value of the Polyakov loop is always zero in a practical simulation with finite volume. 
In the case of heavy fermions, i.e. $\kappa$ is sufficiently small, the canonical partition function with $N=1$ is in proportion to $\langle \hat{\Omega} \rangle$,
since the leading contribution is the Polyakov loop term in the hopping parameter expansion of $\ln \det M$ with the winding number one, and $\det_1 M$ in the equation of $Z_C (T,1)$, e.g. Eq.~(\ref{eq:detNM}), is proportional to $\hat{\Omega}$.
In Fig.~\ref{fig:disU1}, we plot complex values of the Polyakov loop on each configuration.
The configurations are generated by a heat bath method of $U(1)$ lattice gauge theory \cite{Creutz:1983ev} with the standard plaquette gauge action and no dynamical fermions. 
The left panel shows the result of the confinement (symmetric) phase at $\beta=0.90$, and the right panel shows the result of the deconfinement (broken) phase at $\beta=1.10$.
The temporal lattice size $N_t$ is fixed to be 4. 
We adopt four spatial lattice sizes $N_s$. 
Purple, blue, green and red symbols are the results of $N_s=12$, 16, 24 and 32, respectively.
In the confinement phase, the Polyakov loop distribute around $\Omega=0$. 
As the volume increases, the width of the distribution decreases.
In the deconfinement phase, the Polyakov loop is distributed on a circle and there is no volume dependence in the distribution.

The Polyakov loop is an order parameter of the deconfinement phase transition.
In the deconfinement phase, the center symmetry is spontaneously broken and 
$\langle \hat{\Omega} \rangle$ should be nonzero.
However, because of the center symmetry, the probability distribution is symmetric under the $U(1)$ transformation, $\hat{\Omega} \to e^{i \eta} \hat{\Omega}$ for an arbitrary real number $\eta$, as shown in Fig.~\ref{fig:disU1}.
Therefore, the expectation value of $\hat{\Omega}$ is always zero even in the broken phase.
Because the symmetry is not broken in an simulation, to discuss spontaneous symmetry breaking, it is required to break the center symmetry adding an explicit breaking term in the action.
Then, the breaking-term dependence and spatial-volume dependence are investigated.
If $\langle \hat{\Omega} \rangle$ is nonzero in the double limit of zero breaking term and infinite volume, we identify that spontaneous symmetry breaking has occurred.

\begin{figure}[tb]
\begin{center}
\vspace{0mm}
\includegraphics[width=7.8cm]{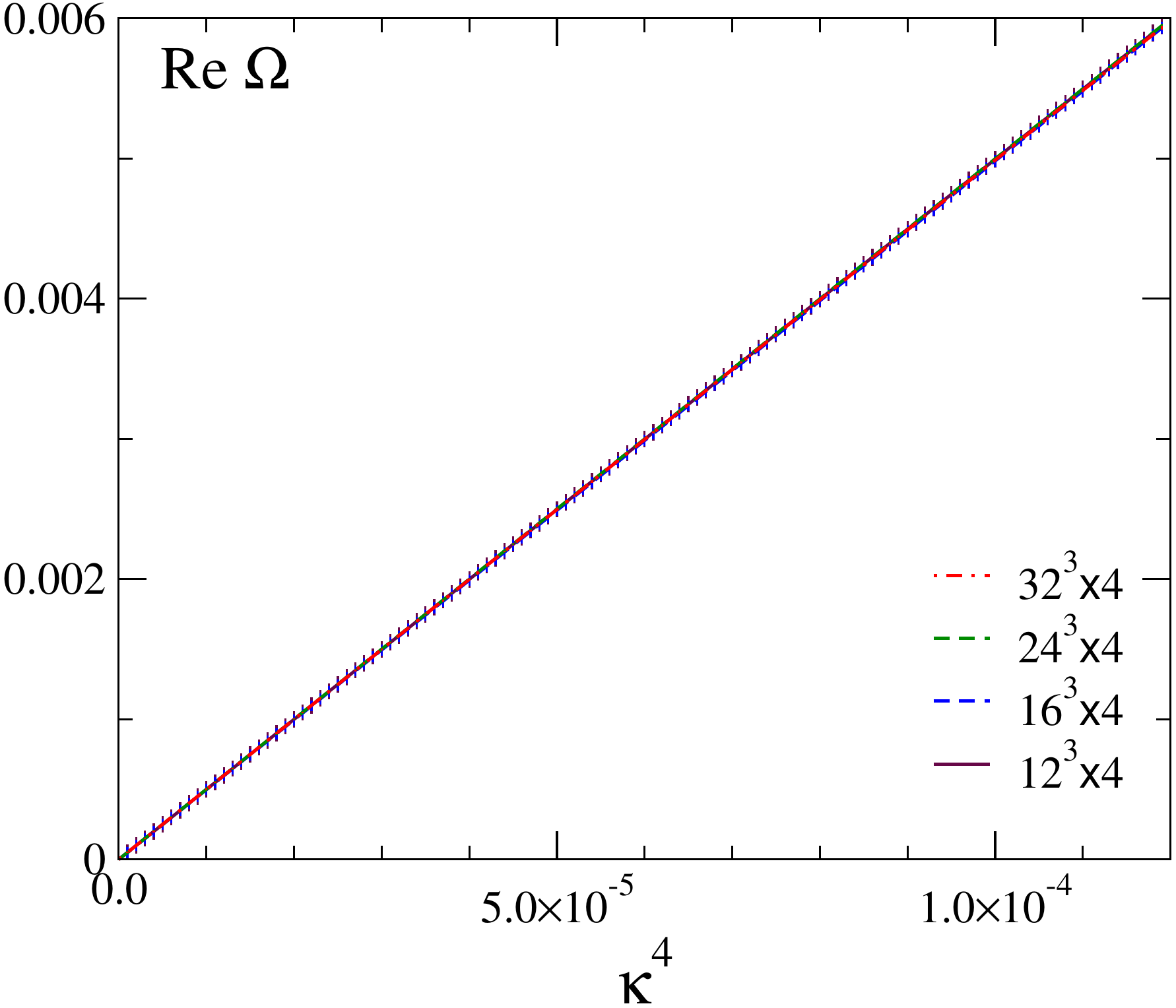}
\hspace{2mm}
\includegraphics[width=7.5cm]{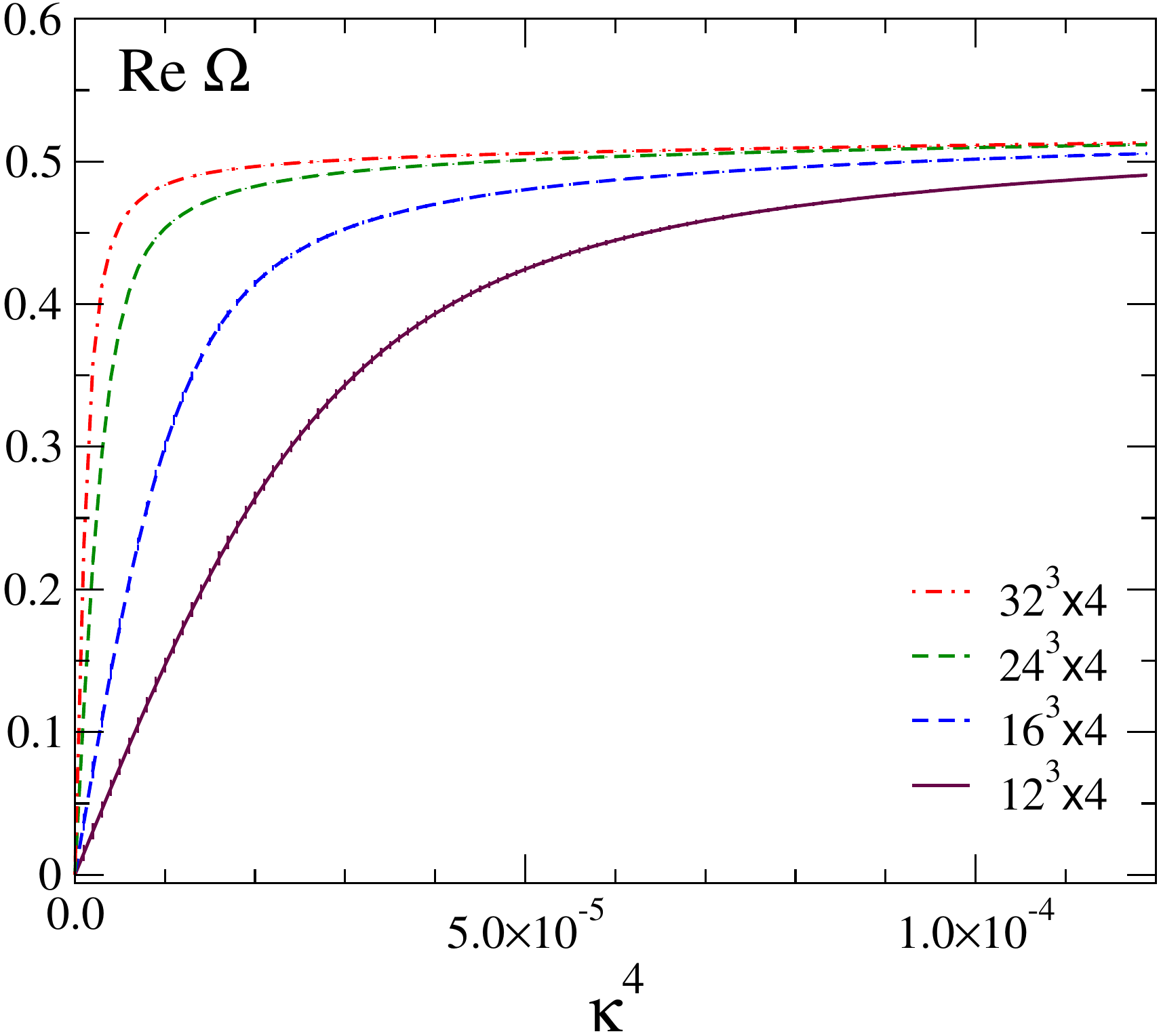}
\vspace{0mm}
\end{center}
\caption{Expectation value of the Polyakov loop as a function of the hopping parameter $\kappa$ measured at $\beta=0.90$ (left) and 1.10 (right) on $N_s^3 \times 4$ lattices with $N_s= 12$ (purple), 16 (blue), 24 (green), and 32 (red).}
\label{fig:rplk4}
\end{figure}

To break the center symmetry, we add one flavor of heavy dynamical fermion.
We approximate the fermion determinant with the leading-order term of the hopping parameter $\kappa$ expansion given in Eq.~(\ref{eq:proddetM}), since we investigate the limit of $\kappa \to 0$.
The plaquette term can be absorbed into the gauge action by shifting 
$\beta \to \beta^* = \beta +16 \kappa^4$.
The expectation value is computed by the reweighting method with the reweighting factor including the Polyakov-loop term,
\begin{eqnarray}
\langle {\rm Re} \hat{\Omega} \rangle_{(\beta, \kappa)} 
&=& \frac{1}{Z_{GC}} \int {\cal D} U \, {\rm Re} \hat{\Omega} \, \det M \, e^{-S_g(\beta)} 
\nonumber \\
& \approx & \frac{1}{Z_{GC}} \int {\cal D} U \, {\rm Re} \hat{\Omega} \, e^{\epsilon V {\rm Re} \hat{\Omega}} \, e^{-S_g(\beta^*)} 
= \frac{\left\langle {\rm Re} \hat{\Omega} \, e^{\epsilon V {\rm Re} \hat{\Omega}} \right\rangle_{(\beta^*,0)}}{\left\langle e^{\epsilon V {\rm Re} \hat{\Omega}} \right\rangle_{(\beta^*, 0)}} ,
\end{eqnarray}
where $\epsilon = 4 \times 2^{N_t} \kappa^{N_t}$, $V=N_s^3$ and 
$\langle \cdots \rangle_{(\beta^*, 0)}$ means the average over quenched configurations at $\beta^*$.
Using the data of $N_t=4$ in Fig.~\ref{fig:disU1}, the expectation value of ${\rm Re} \hat{\Omega}$ is computed.
The number of updates is 1,000,000 for each $\beta$.
The jackknife error is evaluated adopting an appropriate bin size.
We plot the results in Fig.~\ref{fig:rplk4} as a function of $\kappa^{N_t}$.
The left figure is the result of the confinement (symmetric) phase at $\beta=0.90$. 
The results of $N_s=12$ (purple), 16 (blue), 24 (green) and 32 (red) are plotted. 
No volume dependence is observed for $\langle {\rm Re} \hat{\Omega} \rangle$.
Therefore, it is not necessary to take the volume infinity limit, and  
$\langle {\rm Re} \hat{\Omega} \rangle \sim \kappa^{N_t}$ in the confinement phase (symmetric phase).
Then, $\langle {\rm Re} \hat{\Omega} \rangle =0$ in the limit of $\kappa \to 0$.
On the other hand, the right figure is the result of the deconfinement (broken) phase at $\beta=1.10$.
In the deconfinement phase, the Polyakov loop behaves as $\langle {\rm Re} \hat{\Omega} \rangle \sim V \kappa^{N_t}$.
Of course, if $V$ is fixed and the limit of $\kappa \to 0$ is taken, $\langle {\rm Re} \hat{\Omega} \rangle$ becomes always zero.
However, in the thermodynamic limit, i.e. $V \to \infty$, and $\kappa \to 0$, 
$\langle {\rm Re} \hat{\Omega} \rangle$ is expected to become a nontrivial finite value.\footnote{
In this analysis, we ignored the difference between $\beta$ and $\beta^*$ because the difference is very small. 
To be precise, Figure \ref{fig:rplk4} is the results of $\beta^*=0.90$ and 1.10.
Although $\langle {\rm Re} \hat{\Omega} \rangle$ at $\kappa = 0$ is zero within the statistical error due to the center symmetry, in order to make it exactly zero at $\kappa = 0$, we assumed that the generation probability of $\hat{\Omega}$ and $-\hat{\Omega}$ is the same in the calculation of the expectation value.}

Furthermore, when $\kappa$ is sufficiently small, the expectation value of the Polyakov loop can be evaluated by a Taylor expansion of $\kappa^{N_t}$ assuming the $U(1)$ center symmetry.
Using the distribution function of the Polyakov loop in the complex plane, the expectation value can be calculated as follows, 
\begin{eqnarray}
\langle {\rm Re} \hat{\Omega} \rangle_{(\beta, \kappa)} 
&=& \frac{1}{Z_{GC}} \int {\cal D} U \, {\rm Re} \hat{\Omega} \, e^{\epsilon V {\rm Re} \hat{\Omega}} \, e^{-S_g} 
= \frac{1}{2 \pi} \iint |\Omega| \cos \phi \, e^{\epsilon V |\Omega| \cos \phi} \, W(|\Omega|) \, d \phi \, d |\Omega|
\nonumber \\
&=& \frac{\epsilon V}{2} \int |\Omega|^2 \, W(|\Omega|) \, d |\Omega| + \cdots .
\end{eqnarray}
Here, $W(|\Omega|)$ is the probability distribution when the Polyakov loop is $|\Omega| e^{i \phi}$ in the simulation without dynamical fermions.
This probability distribution function is $U(1)$ symmetric, i.e. independent of $\phi$, and is a function of the absolute value $|\Omega|$, as seen in Fig.~\ref{fig:disU1}.
The distribution function is normalized as $\int_0^{\infty} W(|\Omega|) \, d|\Omega| =1$.
The value of $\langle {\rm Re} \hat{\Omega}^n \rangle$ can be also computed, 
\begin{eqnarray}
\langle {\rm Re} \hat{\Omega}^n \rangle_{(\beta, \kappa)} 
&=& \frac{1}{Z_{GC}} \int {\cal D} U \, {\rm Re} \hat{\Omega}^n \, e^{\epsilon V {\rm Re} \hat{\Omega}} e^{-S_g} 
= \frac{1}{2 \pi} \iint |\Omega|^n \cos (n \phi) \, e^{\epsilon V |\Omega| \cos \phi} W(|\Omega|) \, d \phi \, d |\Omega|
\nonumber \\
&=& \frac{(\epsilon V)^n}{n! \ 2^{n}} \int |\Omega|^{2n} \, W(|\Omega|) \, d |\Omega| + \cdots .
\end{eqnarray}
The complex phase can be removed in the above equations.
In the double limit of $V \to \infty$, and $\kappa \to 0$, the value of $\epsilon V$ cannot be determined.
However, for example, the leading term of the ratio 
$\langle {\rm Re} \hat{\Omega}^4 \rangle / \langle {\rm Re} \hat{\Omega}^2 \rangle^2$
does not depend on $\epsilon V$.
The volume dependence of such a quantity is expected to small.
The explicit breaking term is also important for the calculation of the canonical partition function. 
As discussed below, this method can be applied in computing the derivative of canonical partition function.

\section{Canonical partition function with a saddle point approximation}
\label{sec:saddle}

We calculate the canonical partition function following Ref.~\cite{Ejiri:2008xt}. 
Because the fugacity expansion is basically a Laplace transform, the canonical partition function is obtained by the inverse Laplace transform,
\begin{eqnarray}
{\cal Z}_{\rm C}(T,N) 
= \frac{1}{2 \pi} \int_{-\pi}^{\pi} 
e^{-N (\mu_0/T+i\mu_I/T)} {\cal Z}_{\rm GC}(T, \mu_0+i\mu_I) \ 
d \left( \frac{\mu_I}{T} \right) ,
\label{eq:canonicalP} 
\end{eqnarray}
where $\mu_0$ is arbitrary complex number. 
$\mu_I/T$ is an integral variable introduced to constrain the number of particles $N$. 
Regarding as the imaginary part of $\mu/T$, we extend the $\mu/T$ in Eq.~(\ref{eq:fex}) to a complex value.
This equation is equivalent to performing the complex integral on the complex plane of $z$ along a path parallel to the imaginary axis,
\begin{eqnarray}
Z_C(T,N) = \frac{1}{2\pi} \int Z_{GC}(T,zT)e^{-Nz} dz ,
\label{eq:zci}
\end{eqnarray}
where $z=(\mu_0+i\mu_I)/T$.

We consider the case of degenerate $N_{\rm f}$ flavor QCD for simplicity.
The grand partition function:
\begin{align}
Z_{GC}(T,zT)&=\int\mathcal{D}U(\mathrm{det}M(\kappa, z))^{N_{\rm f}}e^{-S_g} 
\end{align}
can not be computed by Monte Carlo simulations.
Thus, we calculate the following ratio, 
\begin{eqnarray}
\frac{{\cal Z}_{\rm GC}(T, \mu)}{{\cal Z}_{\rm GC}(T,0)}
&=& \frac{1}{{\cal Z}_{\rm GC}} \int {\cal D}U 
\left( \frac{\det M(\kappa, \mu/T)}{\det M(\kappa_0, 0)} \right)^{N_{\rm f}}
(\det M(\kappa_0, 0))^{N_{\rm f}} e^{-S_g} \nonumber \\
&=& \left\langle 
\left( \frac{\det M(\kappa, \mu/T)}{\det M(\kappa_0, 0)} \right)^{N_{\rm f}}
\right\rangle_{(T, \mu=0, \kappa_0)} .
\label{eq:normZGC} 
\end{eqnarray}
$Z_{\rm GC}(T,0)$ is the grand partition function at the simulation point of $\mu=0$.
$\langle \cdots \rangle_{(T, \mu=0, \kappa_0)}$ is the expectation value at $\mu=0$ with the hopping parameter $\kappa_0$. 
Substituting this equation into Eq.~(\ref{eq:zci}),
the canonical partition function becomes 
\begin{eqnarray}
Z_C(T,N) &=& \frac{1}{2\pi} Z_{GC}(T,0) 
\int \left\langle e^{VD(z)}e^{-Nz} \right\rangle_{(T, \mu=0)} dz \nonumber \\
&=& \frac{1}{2\pi} Z_{GC}(T,0) \left\langle
\int e^{V(D(z)-\frac{N}{V}z)} dz \right\rangle_{(T, \mu=0)} ,
\label{eq:zclt}
\end{eqnarray}
where $V=N_s^3$ and 
\begin{eqnarray}
e^{VD(z)} \equiv \left( \frac{\det M(\kappa, z)}{\det M(\kappa_0, 0)} \right)^{N_{\rm f}} .
\label{eq:Dz}
\end{eqnarray}
In the second line of Eq.~(\ref{eq:zclt}), we changed the calculation order of the complex integral of $z$ and the path integral of the gauge field.

Here, the particle number density is expressed as $\rho = N/V$ in lattice units.
In physical units, the density $\rho_{\rm phy}$ is given by $\rho_{\rm phy} / T^3 = \rho N_t^3$. 
For simplicity, we integrate $z$ by a saddle point approximation \cite{Ejiri:2008xt}. 
This approximation is valid for large volumes. 
The saddle point condition is 
\begin{eqnarray}
\frac{d}{dz} \left[ D(z) - \rho z \right] = \frac{d D(z)}{dz} - \rho =0 .
\end{eqnarray}
Let $z_0=x_0+iy_0$ be the saddle point where $z$ satisfies the saddle point condition. 
\begin{eqnarray}
{\cal Z}_{\rm C}(T, \rho V) &=& \frac{1}{2 \pi} {\cal Z}_{\rm GC}(T,0) 
\left\langle \int_{-\pi}^{\pi} 
e^{-N (z_0+ix)} \left( \frac{\det M (\kappa, z_0+ix)}{\det M(\kappa_0, 0)} \right)^{N_{\rm f}}
dx \right\rangle_{(T, \mu=0, \kappa_0)} \nonumber \\
&=& \frac{1}{2 \pi} {\cal Z}_{\rm GC}(T,0) \left\langle \int_{-\pi}^{\pi} 
\exp \left[ V \left(D (z_0) - \rho z_0 
- \frac{1}{2} D'' (z_0) x^2 + \cdots \right) \right] 
dx \right\rangle_{(T, \mu=0, \kappa_0)} \nonumber \\
& \approx & \frac{1}{\sqrt{2 \pi}} {\cal Z}_{\rm GC} (T,0)
\left\langle \exp \left[ V \left( D(z_0) - \rho z_0 \right) \right] 
e^{-i \alpha/2} \sqrt{ \frac{1}{V |D''(z_0)|}}
\right\rangle_{(T, \mu=0, \kappa_0)} ,
\label{eq:zcspa}
\end{eqnarray}
where $ D''(z) = d^2 D(z) /dz^2$ and $D''(z)=|D''(z)| e^{i \alpha}$. 
The derivative with respect to $\rho$ is obtained by the following equation,
\begin{eqnarray}
%- \frac{\Delta \ln {\cal Z}_{\rm C} (T,N) }{\Delta N} & = & 
- \frac{1}{V} \frac{\partial \ln {\cal Z}_{\rm C} (T, \rho V)}
{\partial \rho} 
\approx \frac{
\left\langle z_0 \ \exp \left[ V \left( D(z_0)
- \rho z_0 \right) \right] 
e^{-i \alpha /2} \sqrt{ \frac{1}{V |D''(z_0)|}}
\right\rangle_{(T, \mu=0, \kappa_0)}}{
\left\langle \exp \left[ V \left( D(z_0) 
- \rho z_0 \right) \right] 
e^{-i \alpha /2} \sqrt{ \frac{1}{V |D''(z_0)|}}
\right\rangle_{(T, \mu=0, \kappa_0)}} .
\label{eq:chemap}
\end{eqnarray}
This equation can be calculated by the Monte Carlo method.

\section{$U(1)$ lattice gauge theory with heavy fermions}
\label{sec:u1gt}

\subsection{Canonical partition function of heavy fermions}

We demonstrate the calculation of the canonical partition function in $U(1)$ lattice gauge theory with degenerate $N_{\rm f}$ flavors of heavy dynamical fermions. 
$D(z)$ in Eq.~(\ref{eq:zclt}) is evaluated by the hopping parameter expansion, 
\begin{eqnarray}
D(z) = 96 N_t N_{\rm f} \kappa^4 \hat{P} + 2 \times 2^{N_t} N_{\rm f} \kappa^{N_t} \left[ e^z \hat{\Omega} + e^{-z} \hat{\Omega}^{\dag} \right] + \cdots .
\end{eqnarray}
Assuming that $\kappa$ is sufficiently small, we approximate $D(z)$ by the terms $\hat{P}$ and $\hat{\Omega}$.
The first and second derivatives of $D(z)$ with respect to $z$ are 
\begin{eqnarray}
\frac{\partial D(z)}{\partial z} &=& 2 \times 2^{N_t} N_{\rm f} \kappa^{N_t} \left[ e^z \hat{\Omega} -e^{-z} \hat{\Omega}^{\dag} \right] + \cdots , \\
\frac{\partial ^2 D(z)}{\partial z^2} &=& 2 \times 2^{N_t} N_{\rm f} \kappa^{N_t} \left[ e^z \hat{\Omega} + e^{-z} \hat{\Omega}^{\dag} \right] + \cdots .
\end{eqnarray}
The saddle point $z_0=x_0+iy_0$ is defined as 
\begin{eqnarray}
\left[ \frac{\partial D(z)}{\partial z} - \rho \right]_{z=z_0} = 0.
\end{eqnarray}
Thus, 
\begin{eqnarray}
2 \times 2^{N_t} N_{\rm f} \kappa^{N_t} \left( e^{z_0} \hat{\Omega} -e^{-z_0} \hat{\Omega}^{\dag} \right) = \rho .
\label{eq:saddle}
\end{eqnarray}
Because $\rho$ is a real number, 
the complex phase of $(e^{iy_0} \hat{\Omega} -e^{-iy_0} \hat{\Omega}^{\dag})$ is zero. 
The imaginary part of saddle point $y_0$ is determined as
\begin{eqnarray}
\tan y_0 = -\frac{\mathrm{Im} \hat{\Omega}}{\mathrm{Re} \hat{\Omega}}  
\hspace{5mm} {\rm or} \hspace{5mm}  
y_0 = - \arctan \left( \frac{\mathrm{Im} \hat{\Omega}}{\mathrm{Re} \hat{\Omega}} \right) 
=- \mathrm{Arg} \hat{\Omega} .
\end{eqnarray}

Substituting $y_0$ into Eq.~(\ref{eq:saddle}),
\begin{eqnarray}
2 \times 2^{N_t} N_{\rm f} \kappa^{N_t} \left[ e^{x_0} |\hat{\Omega}| -e^{-x_0} |\hat{\Omega}| \right] 
=4 \times 2^{N_t} N_{\rm f} \kappa^{N_t} |\hat{\Omega}| \sinh x_0 
= \rho .
\end{eqnarray}
Solving $x_0$, 
\begin{eqnarray}
x_0  &=& \mathrm{arcsinh} \left( \frac{\rho}{4 \times 2^{N_t} N_{\rm f} \kappa^{N_t} |\hat{\Omega}|} \right) \nonumber \\
&=& \ln \left( \frac{\rho} {4\times 2^{N_t}N_{\rm f} \kappa^{N_t} |\hat{\Omega}|} 
+ \sqrt{ \left(\frac{\rho} {4\times 2^{N_t}N_{\rm f} \kappa^{N_t} |\hat{\Omega}|} \right)^2+1} \right) .
\label{eq:x0}
\end{eqnarray}
The second derivative of $D(z)$ at $z=z_0$, $D''(z_0)$, is given by 
\begin{eqnarray}
D''(z_0)
=2 \times 2^{N_t} N_{\rm f} \kappa^{N_t} \left[ e^{x_0} |\hat{\Omega}| + e^{-x_0} |\hat{\Omega}| \right]
=4 \times 2^{N_t} N_{\rm f} \kappa^{N_t} |\hat{\Omega}| \cosh x_0 .
\label{eq:D2z0}
\end{eqnarray}
Similarly, $D(z)$ at $z=z_0$ is
\begin{eqnarray}
D(z_0) = 96 N_t N_{\rm f} \kappa^4\hat{P} + 4\times2^{N_t} N_{\rm f} \kappa^{N_t} |\hat{\Omega}| \cosh x_0 .
\label{eq:Dz0}
\end{eqnarray}
In this approximation, $x_0$, $D(z_0)$, and $D''(z_0)$ are real positive numbers.
We substitute these equations into Eq.~(\ref{eq:chemap}). 
The derivative of $Z_C$ is 
\begin{eqnarray}
-\frac{1}{V} \frac{\partial \mathrm{ln} Z_C(T,N)} {\partial \rho} 
\approx \frac{\left\langle z_0 \exp (F+i\theta) \right\rangle} {
\left\langle \exp (F+i\theta) \right\rangle}, 
\end{eqnarray}
where $F$ and $\theta$ are defined as
\begin{eqnarray}
F &=& V(D(z_0) - \rho x_0) -\frac{1}{2} \ln [V D^{\prime\prime}(z_0)] ,
\label{eq:f} \\
\theta &=& -V \rho \, \mathrm{Arg} \hat{\Omega} 
= -V \rho \, \arctan \left( \frac{{\rm Im} \hat{\Omega}}{{\rm Re} \hat{\Omega}} \right) .
\label{eq:theta}
\end{eqnarray}

The Monte Carlo simulation is performed without dynamical fermions, $\kappa_0=0$ and $\det M(0, 0)=1$, and $\langle \cdots \rangle$ means the expectation value of the quenched simulations.
The term proportional to $\hat{P}$ in Eq.~(\ref{eq:Dz0}) can be absorbed into the gauge action by shifting 
$\beta \to \beta^* = \beta +16 N_{\rm f} \kappa^4$ in the quenched simulation, and the shift of $\beta$ is very small in this calculation.
Therefore, we omit Wilson-loop-type terms such as the plaquette term and concentrate on the effect of the Polyakov-loop term.
Then, $x_0$ and $F$ are real functions of $\rho$ and the absolute value of $\Omega$. 
(See Eqs.~(\ref{eq:x0}), (\ref{eq:D2z0}) and (\ref{eq:Dz0}).)
On the other hand, $\theta$ is a real function of $\rho$ and the complex phase of $\Omega$. 

We calculate the derivative of the canonical partition function by classifying the configurations by the value of $|\Omega|$ in the Monte Carlo simulation.
\begin{eqnarray}
-\frac{1}{V} \frac{\partial \ln Z_C(T, V \rho)} {\partial \rho} 
&\approx& \frac{\int \langle z_0 \, \exp [F+i\theta] \rangle_{|\Omega|} \, W(|\Omega|) \, d|\Omega|
} {\int \langle \exp [F+i\theta] \rangle_{|\Omega|} \, W(|\Omega|) \, d|\Omega|} \nonumber \\
&=& \frac{\int (x_0 \langle \cos \theta \rangle_{|\Omega|} +\langle y_0 \sin \theta \rangle_{|\Omega|}) \, e^F \, W(|\Omega|) \, d|\Omega|
}{\int e^F \langle \cos \theta \rangle_{|\Omega|} \, W(|\Omega|) \, d|\Omega|} , 
\label{eq:dzc}
\end{eqnarray}
where $\langle \cdots \rangle_{|\Omega|}$ means that each configuration is classified by the value of $|\Omega|$ and the expectation value is calculated for each value of $|\Omega|$, i.e.  
\begin{eqnarray}
\langle \cdots \rangle_{|\Omega|=x} = \frac{\langle \cdots 
\delta(|\hat{\Omega}| -x) \rangle}{\langle \delta(|\hat{\Omega}| -x) \rangle} ,
\end{eqnarray}
and $W(|\Omega|)$ is the probability distribution function of $|\Omega|$.
However, because $\theta = -N y_0$ and $y_0 = - \mathrm{Arg} \hat{\Omega} \equiv - \phi$, 
$\langle \cos \theta \rangle_{|\Omega|} = \langle \cos N \phi \rangle_{|\Omega|} =0$.
Namely, the phase average $\langle \cos \theta \rangle_{|\Omega|}$ is exactly zero due to the $U(1)$ symmetric distribution function of $\Omega$. 
So to speak, this is the ultimate sign problem.
Therefore, $\partial \ln Z_C / \partial \rho$ is indefinite. 

Here, we neglect a term 
$\int e^F \langle y_0 \sin \theta \rangle_{|\Omega|} W(|\Omega|) \, d|\Omega|$ 
in Eq.~(\ref{eq:dzc}).
Because of the $U(1)$ center symmetry, this can be computed as follows:
\begin{eqnarray}
\langle y_0 \sin \theta \rangle_{|\Omega|} 
&=& \frac{1}{2 \pi} \int_{c}^{c+2\pi} \phi \sin (N \phi) \, d \phi
= \frac{1}{2 \pi} \left[ -\frac{\phi}{N} \cos (N \phi) 
  +\frac{1}{N^2} \sin (N \phi) \right]_{c}^{c+2\pi}
\nonumber \\
&=& -\frac{1}{N} \cos (N c) .
\end{eqnarray}
This value of $\langle y_0 \sin \theta \rangle_{|\Omega|}$ changes if the upper and lower bounds of the integration are changed while the integration range remains $2 \pi$.
We thus define this quantity as the product of an appropriate convergence factor, which satisfies 
$\lim_{\phi \to \infty} e^{-\xi \phi} =0$. Then, this term vanishes,
\begin{eqnarray}
\langle y_0 \sin \theta \rangle_{|\Omega|} 
&\equiv& \lim_{\xi \to 0, \ m \to \infty} \frac{1}{2m \pi} \int_0^{2m \pi} \phi \sin (N \phi) \, e^{-\xi \phi} \, d \phi
\nonumber \\ 
&=& \lim_{\xi \to 0, \ m \to \infty} \frac{1}{2m \pi} \left[ -\frac{e^{-\xi \phi}}{(N^2 +\xi^2)^2} \left\{ \left( N(N^2 +\xi^2) \phi +2N \xi \right) \cos (N \phi) 
\right. \right. \nonumber \\ && \hspace{20mm} \left. \left.
+\left( \xi (N^2 +\xi^2) \phi -N^2 +\xi^2 \right) \sin (N \phi) \right\} \right]_0^{2m \pi} 
\nonumber \\
%&=& \lim_{\xi \to 0, \ m \to \infty} \frac{-e^{-2m \pi \xi} (2m \pi N(N^2 +\xi^2) + 2N \xi) 
%+2N \xi}{2m \pi (N^2 +\xi^2)^2}
&=& \lim_{\xi \to 0, \ m \to \infty} \left[ \frac{-N e^{-2m \pi \xi}}{N^2 +\xi^2}
+\frac{N \xi (1-e^{-2m \pi \xi})}{m \pi (N^2 +\xi^2)^2} \right]
=0,
\end{eqnarray}
where $\xi$ is a real number and $m$ is an integer number.
Moreover, $\langle y_0 \sin \theta \cos^n \phi \rangle_{|\Omega|} $ also vanishes for any positive integer $n$, since it is given by the sum of  
$\int \phi \sin (k \phi) \, d \phi$ with an appropriate integer $k$.
When we break the center symmetry adding a heavy dynamical fermion, 
the calculation of terms like $\langle y_0 \sin \theta \cos^n \phi \rangle_{|\Omega|} $ is required, but such terms do not contribute to Eq.~(\ref{eq:dzc}).
Therefore, we neglect the term of $\langle y_0 \sin \theta \rangle_{|\Omega|}$.

%We define an effective potential $U_{\rm eff}$, 
%\begin{eqnarray}
%U_{\rm eff}=-\ln w -F-\ln \langle \cos \theta \rangle_{|\Omega|} .
%\end{eqnarray}

We use the method explained in Sec.~\ref{sec:centerC} to compute $\langle \cos \theta \rangle_{|\Omega|}$ avoiding the sign problem.
An additional heavy fermion with a small hopping parameter $\kappa_h$ is introduced to break the center symmetry.
The fermion determinant is approximated by $e^{\epsilon V {\rm Re} \hat{\Omega}}$, 
where $\hat{\Omega}$ is the Polyakov-loop operator and $\epsilon = 4 \times 2^{N_t} \kappa_h^{N_t}$, and the Wilson-loop-type terms are disregarded.
When $\epsilon$ is small, the denominator of Eq.~(\ref{eq:dzc}) is computed as follows,
\begin{eqnarray}
\langle e^F \cos \theta \rangle
&=& \frac{1}{Z} 
\int {\cal D} U e^F \cos \theta \, e^{\epsilon V {\rm Re} \hat{\Omega}} e^{-S_g} 
= \frac{1}{Z} 
\int {\cal D} U e^F \cos (N \phi) \, e^{\epsilon V |\hat{\Omega}| \cos \phi} e^{-S_g} 
\nonumber \\
&=& \frac{1}{2 \pi} \iint e^F \cos (N \phi) \, e^{\epsilon V |\Omega| \cos \phi} W(|\Omega|) \, d \phi \, d |\Omega|
\nonumber \\
&=& \int e^F W(|\Omega|) \, \frac{1}{2^N N!} \, (\epsilon V)^N \, |\Omega|^N \, 
d |\Omega| + \cdots .
\end{eqnarray}
Similarly, the numerator of Eq.~(\ref{eq:dzc})
\begin{eqnarray}
\langle x_0 \, e^F \cos \theta \rangle 
= \int x_0 \, e^F \, W(|\Omega|) \, \frac{1}{2^N N!} \, (\epsilon V)^N |\Omega|^N \, d| \Omega | + \cdots .
\end{eqnarray}
Thus, the derivative of $Z_C$ is obtained by
\begin{eqnarray}
-\frac{1}{V} \frac{\partial \ln Z_C(T,V \rho)}{\partial \rho} 
\approx 
\frac {\int x_0 \, e^F \, W(|\Omega|) \, \frac{1}{2^N N!} \, (\epsilon V)^N |\Omega|^N \, d|\Omega|}{\int e^F \, W(|\Omega|) \, \frac{1}{2^N N!} \, (\epsilon V)^N |\Omega|^N \, d|\Omega| }
=\frac {\int x_0 \, e^{-U_{\rm eff}} \, d|\Omega| } {\int  e^{-U_{\rm eff}} \, d|\Omega| } ,
\label{eq:dzcu}
\end{eqnarray}
where the effective potential $U_{\rm eff} (|\Omega|)$ is defined as 
\begin{eqnarray}
U_{\rm eff} (|\Omega|) = -\ln W(|\Omega|) -F -N \ln |\Omega|.
\label{eq:Ueff}
\end{eqnarray}
Here, we should note that $\langle e^F \cos \theta \rangle$ and $\langle x_0 \, e^F \cos \theta \rangle$ go to zero in the limit of $\epsilon \to 0$, causing the sign problem.
However, the factors $(\epsilon V)^N$ in the numerator and denominator of Eq.~(\ref{eq:dzcu}) are canceled, and this equation becomes calculable without considering the additional factor $\epsilon V$. 
We moreover find that this quantity is approximately equal to the value of $x_0$ at the $|\Omega|$ where the effective potential $U_{\rm eff} (|\Omega|)$ is the minimum in the case that the volume is sufficiently large.

\subsection{Numerical simulations}

\begin{figure}[tb]
\begin{minipage}{0.47\hsize}
\begin{center}
\vspace{0mm}
\includegraphics[width=7.4cm]{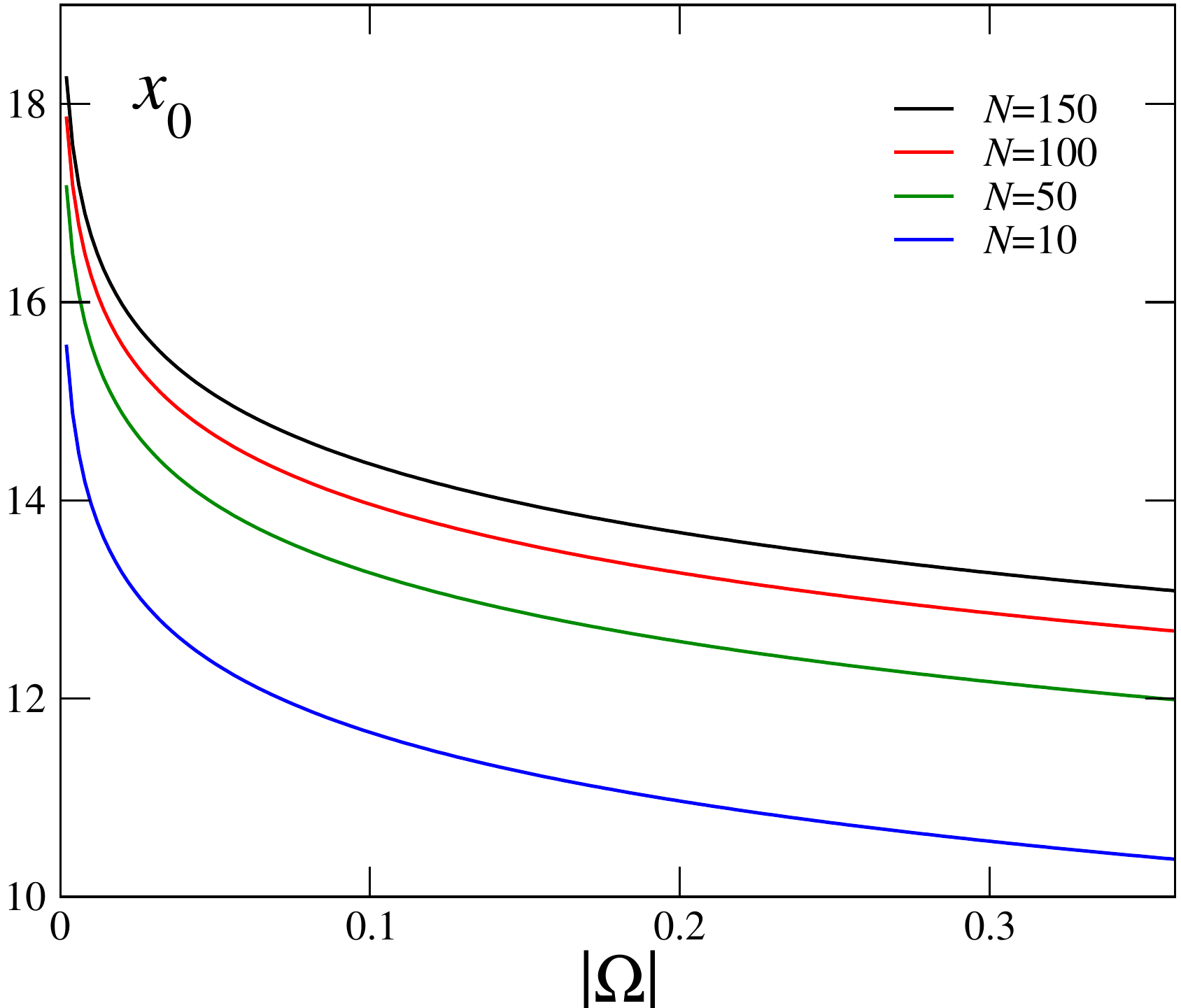}
\vspace{0mm}
\end{center}
\caption{Real part of the saddle point $x_0$ as a function of $|\Omega|$ for $N=10$ (blue), 50 (green), 100 (red), and 150 (black).}
\label{fig:x0}
\end{minipage}
\hspace{1mm}
\begin{minipage}{0.47\hsize}
\begin{center}
\vspace{0mm}
\includegraphics[width=7.6cm]{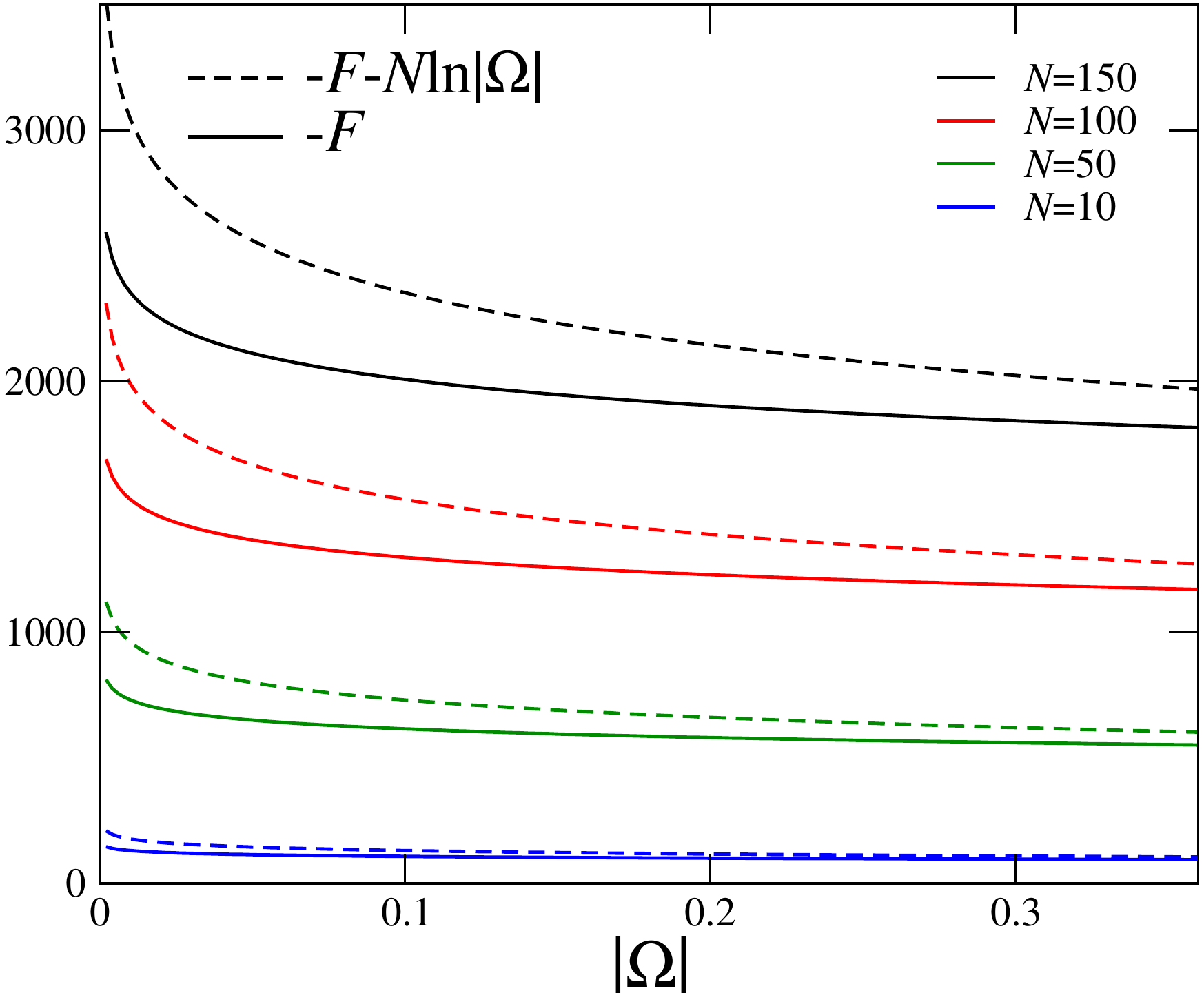}
\vspace{0mm}
\end{center}
\caption{$-F$ (solid line) and $-F-N \ln |\Omega|$ (dashed line) 
as a function of $|\Omega|$ for $N=10$ (blue), 50 (green), 100 (red), and 150 (black).}
\label{fig:F}
\end{minipage}
\end{figure}

We perform Monte Carlo simulations of $U(1)$ lattice gauge theory with the standard Wilson action at several inverse gauge couplings $\beta=1/g^2$ near the transition point $\beta_c$. 
The effective potential $U_{\rm eff} (|\Omega|)$ and saddle point are numerically calculated, and the effective potential $V_{\rm eff}(N)$ for the particle number $N$ are investigated.
The lattice size is $N_s^3 \times N_t= 24^3 \times 6$.
We adopt $N_{\rm f}=2$ and $\kappa =0.025$.
Since the chiral limit of free Wilson fermion is $\kappa_c=1/8$, the $\kappa$ we adopt is small. 
The real part of saddle point $x_0$ for each $|\Omega|$ and $N= \rho V$ is given by Eq.~(\ref{eq:x0}), which is plotted in Fig.~\ref{fig:x0}.
Also, $F$ of Eq.~(\ref{eq:f}) is calculated from only $|\Omega|$ and $\rho$. 
The solid lines in Fig.~\ref{fig:F} are $F$ and the dashed lines are $F - N \ln |\Omega|$ for each $|\Omega|$ and $N$.
The blue, green, red, and black curves are the results of $N=10$, 50, 100, and 150, respectively.

Using a pseudo heat bath algorithm \cite{Creutz:1983ev}, the configurations are generated at thirteen $\beta$ values:  
%in the range from $\beta=1.0000$ to $1.0240$.
$\beta= 1.000, 1.004, 1.006, 1.008, 1.009, 1.0094, 1.0096, 1.010, 1.012, 1.014, 1.016, 1.020,$ and $1.0240$.
The data are taken until there are 1,000,000 heat bath sweeps at each $\beta$.
The multipoint reweighting method \cite{Ferrenberg:1989ui,Iwami:2015eba} are used to combine the data generated at different $\beta$.
The statistical errors are estimated by the jackknife method with the bin size chosen such that the errors are saturated.

\begin{figure}[tb]
\begin{center}
\vspace{0mm}
\includegraphics[width=7.5cm]{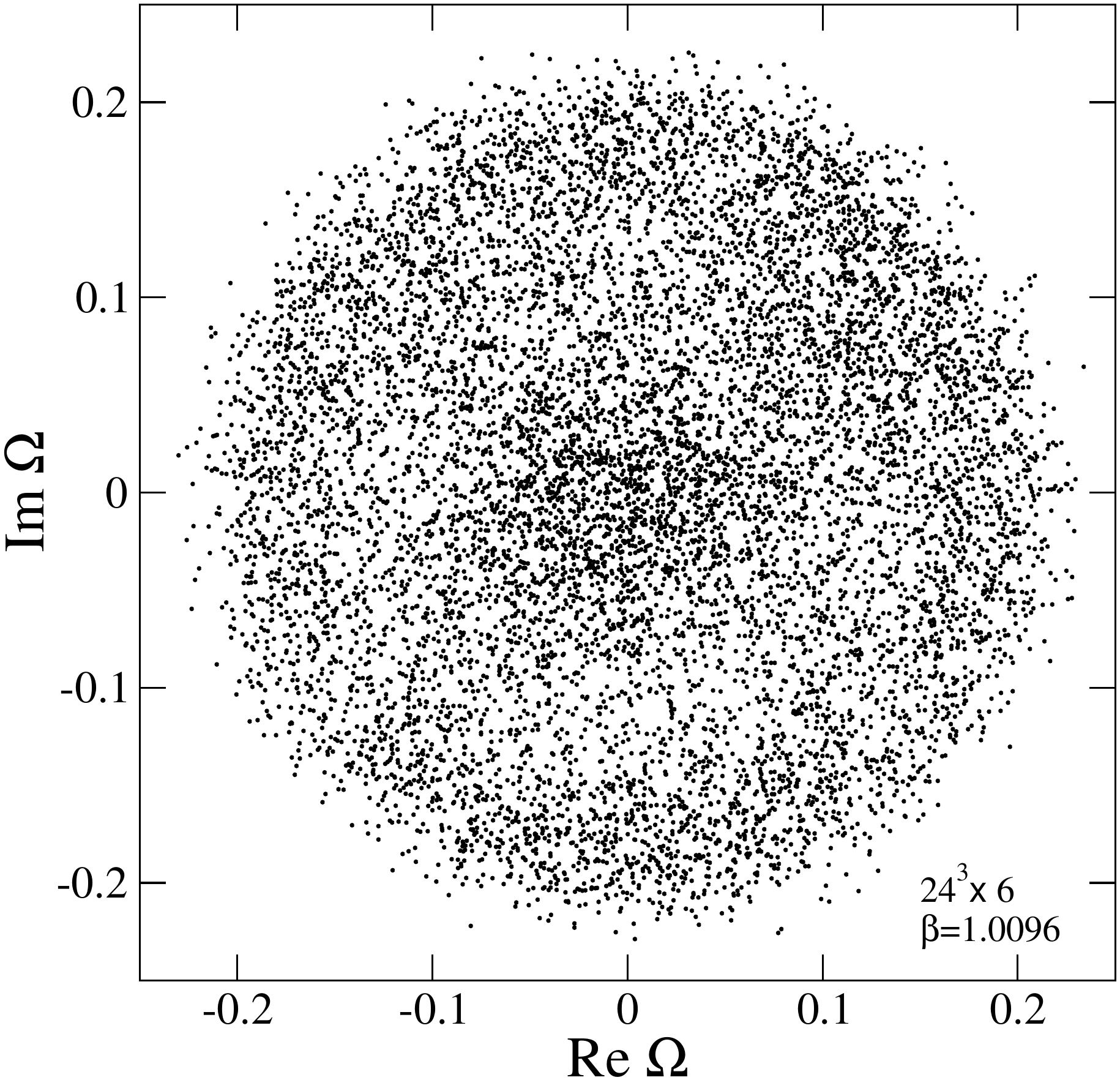}
\vspace{0mm}
\end{center}
\caption{Polyakov-loop distribution at the transition point $\beta=1.0096$ on the $24^3 \times 6$ lattice.}
\label{fig:disbc}
\end{figure}

\begin{figure}[tb]
\begin{minipage}{0.47\hsize}
\begin{center}
\vspace{0mm}
\includegraphics[width=7.7cm]{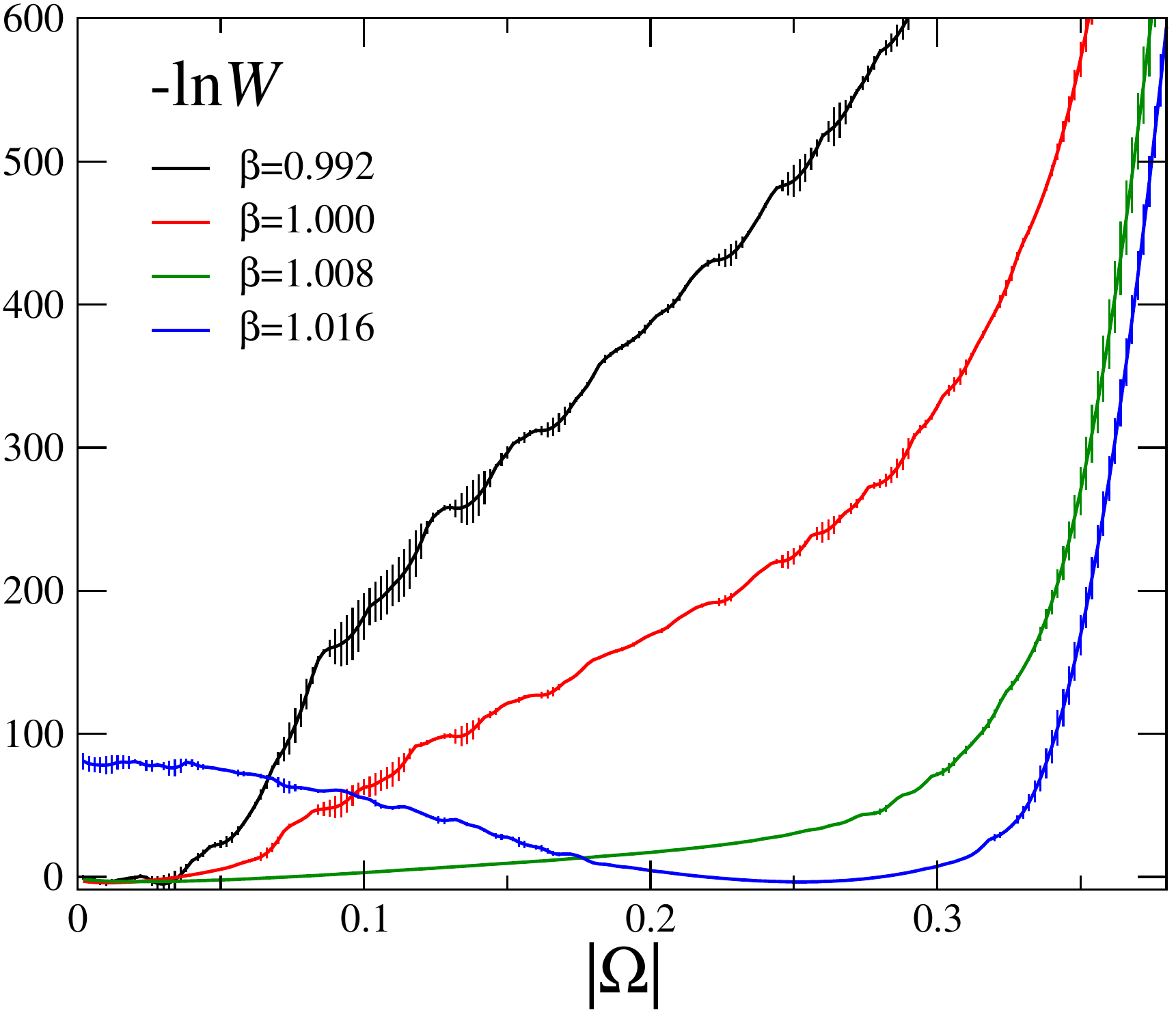}
\vspace{0mm}
\end{center}
\caption{$- \ln W(|\Omega|)$ as a function of $|\Omega|$ for $\beta=0.992$ (black), 1.000 (red), 1.008 (green), and 1.016 (blue).}
\label{fig:wo}
\end{minipage}
\hspace{1mm}
\begin{minipage}{0.47\hsize}
\begin{center}
\vspace{0mm}
\includegraphics[width=7.7cm]{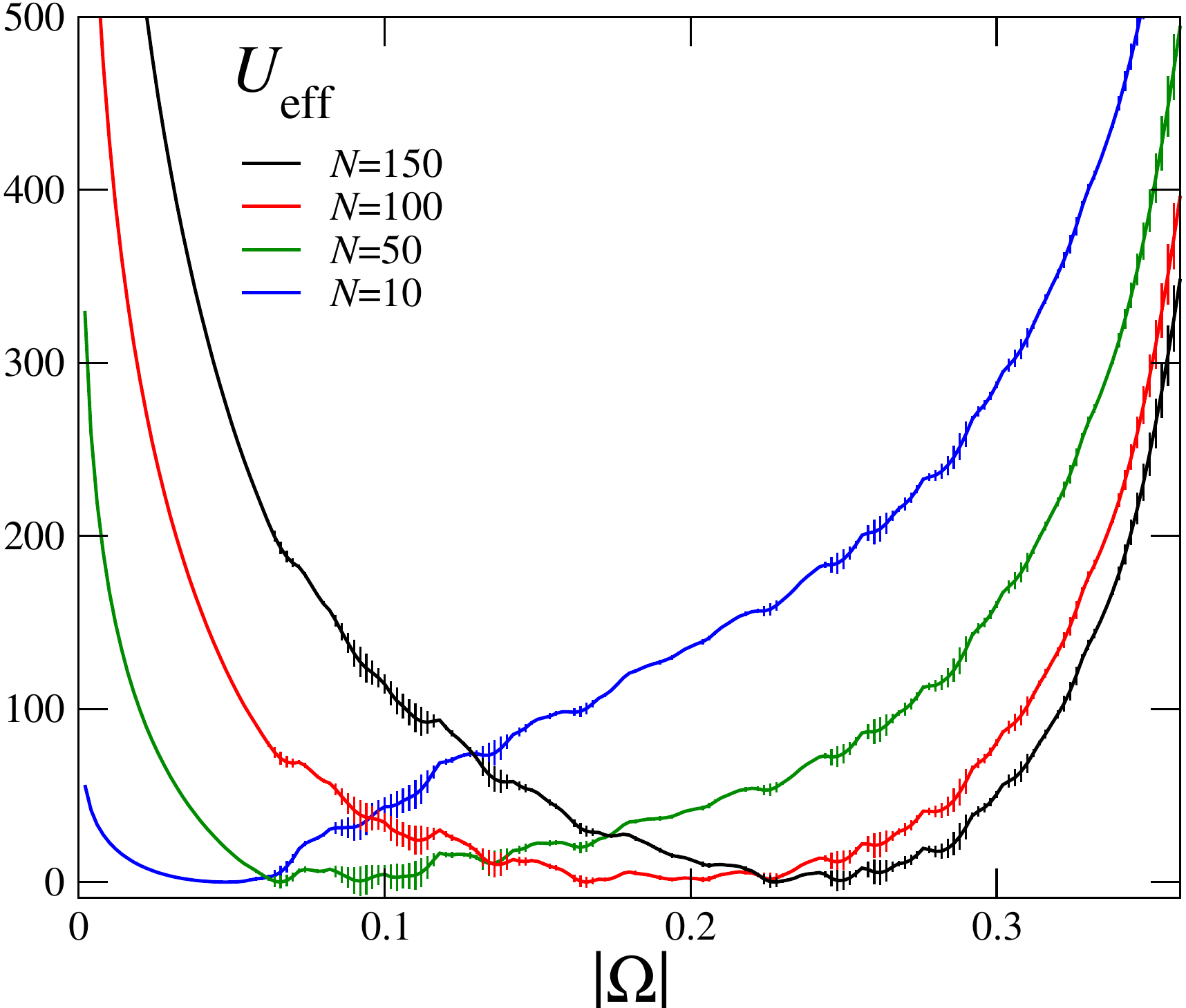}
\vspace{0mm}
\end{center}
\caption{Effective potential $U_{\rm eff} (|\Omega|)$ defined in Eq.~(\ref{eq:Ueff}) at $\beta=1.00$ for $N=10$ (blue), 50 (green), 100 (red), and 150 (black).}
\label{fig:epot}
\end{minipage}
\end{figure}

Figure \ref{fig:disbc} shows the distribution of the Polyakov loop for each configuration at $\beta=1.0096$. The horizontal and vertical axes are the real part and imaginary part, respectively. 
This figure indicates that the phase transition at $\beta=1.0096$ is of very weak first-order transition, where two phases coexist.
The distributions near the origin are the configurations of the confinement phase.
On the other hand, the points around the circle of $|\Omega|=0.15$ are those of the deconfinement phase.

To compute the probability distribution function of $|\Omega|$, $W(|\Omega|)$, we approximate the delta function with a Gaussian function like
\begin{eqnarray}
\delta (x) \approx \frac{1}{\Delta \sqrt{\pi}} \exp \left[ -\left( \frac{x}{\Delta} \right)^2 \right] .
\label{eq:deltaap}
\end{eqnarray}
We choose the width parameter $\Delta$ by considering a balance between the resolution of the distribution function and its statistical error. 
The values of $\Delta$ we adopt are 0.0025.
The result of $-\ln W(|\Omega|)$ is shown in Fig.~\ref{fig:wo} for 
$\beta=0.992$ (black), 1.000 (red), 1.008 (green), and 1.016 (blue).
The results of the effective potential $U_{\rm eff} (|\Omega|)$ in Eq.~(\ref{eq:dzcu}) at $\beta=1.00$ is plotted in Fig.~\ref{fig:epot} for $N=10$ (blue), 50 (green), 100 (red), and 150 (black), as an example.
Here, since a constant may be added to the effective potential, in Figs.~\ref{fig:wo} and \ref{fig:epot}, the constants are added so that the minimum values of $-\ln W(|\Omega|)$ and $U_{\rm eff} (|\Omega|)$ become zero.

\begin{figure}[tb]
\begin{center}
\vspace{0mm}
\includegraphics[width=9.5cm]{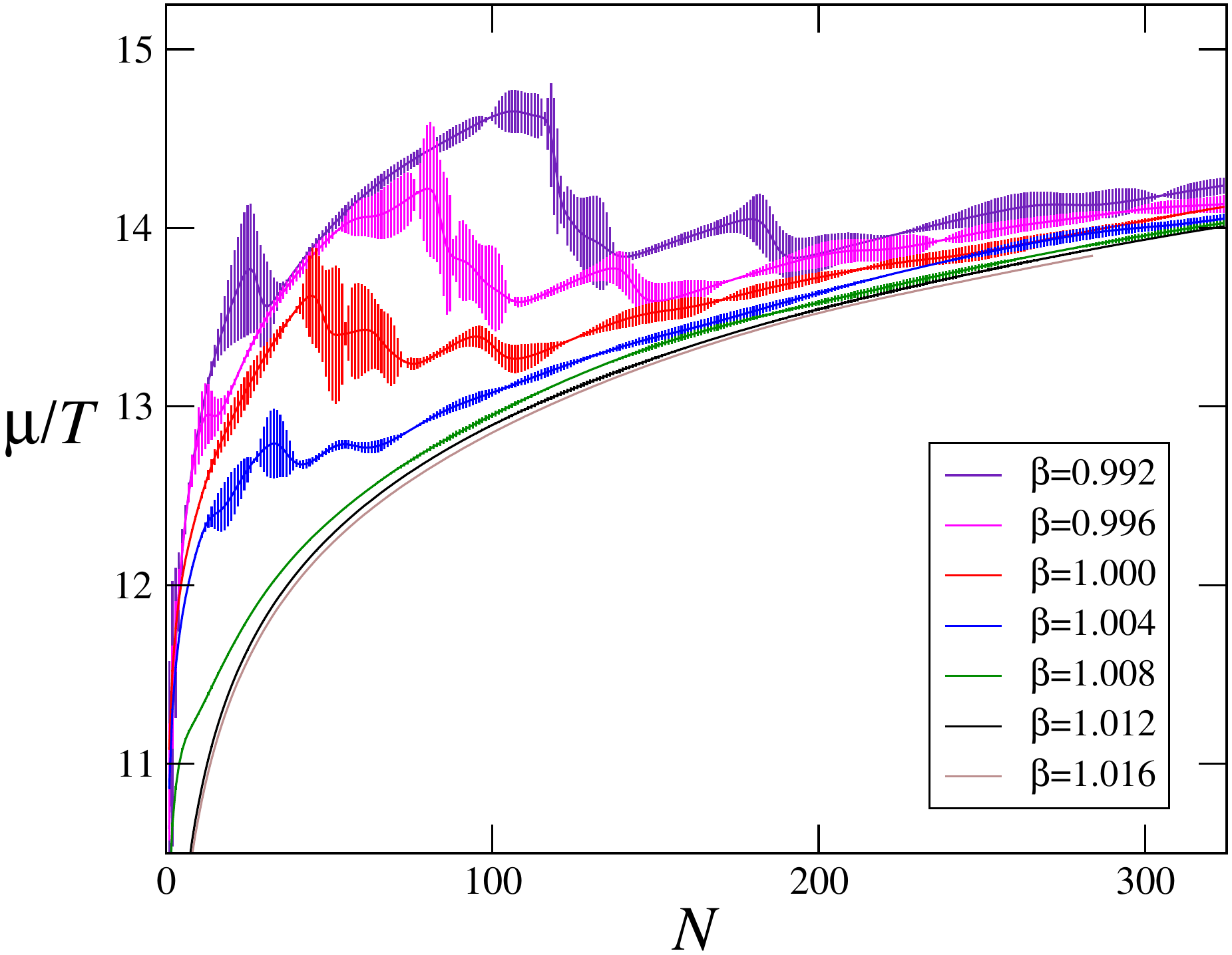}
\vspace{0mm}
\end{center}
\caption{Chemical potential $\mu /T$ for which the number of particles with the maximum generation probability is $N$ at $\kappa =0.025$ for each $\beta$ in $U(1)$ lattice gauge theory with two flavors.}
\label{fig:muvsn}
\end{figure}

We then calculate the derivative of the canonical partition function. 
We use the following equation to remove $W(|\Omega|)$ from Eq.~(\ref{eq:dzcu}) 
so that the equation is independent of the approximation of the delta function, Eq.~(\ref{eq:deltaap}), 
\begin{eqnarray}
-\frac{1}{V} \frac{\partial \ln Z_C (T,V \rho)}{\partial \rho} =
\frac {\int {\cal D} U \, x_0 \, e^F |\hat{\Omega}|^N}{\int {\cal D} U \, e^F |\hat{\Omega}|^N}
\end{eqnarray}
This quantity is $\mu/T$ where the density with the maximum generation probability is $\rho$ and is computable without the sign problem.
We plot the result as a function of $N= V \rho$ for $\beta = 0.992$ (purple), 0.996 (magenta), 1.00 (red), 1.004 (blue), 1.008 (green), 1.0012 (black), and 1.016 (brown) in Fig.~\ref{fig:muvsn}.
The fine up and down vibrations in Fig.~\ref{fig:muvsn} may be an artifact caused by the fact that the effective potential in Fig.~\ref{fig:epot} is not smooth, since this quantity is the value of $x_0$ at the $|\Omega|$ where $U_{\rm eff}$ is the minimum.
Ignoring the fine vibrations, when $\beta$ is in the deconfinement phase at zero density, the chemical potential is monotonically increasing.
However, in the case of a confined phase at zero density, as the density increases, the chemical potential drops once and increases again.
This means that there are multiple $N$s for a specific $\mu/T$.
This is a typical sign when crossing the first-order phase transition.

\subsection{Higher-order terms of the hopping parameter expansion}

As seen in the previous section, our method to solve the problem of the center symmetry and the sign problem works well in $U(1)$ gauge theory.
In the above calculation, the quark determinant has been evaluated by approximating only the leading-order term of the hopping parameter expansion. 
We extend this method to include higher-order terms of the expansion.

We classify each term of the hopping parameter expansion by winding number and denote $D(z)$ in Eq.~(\ref{eq:Dz}) as follows,
\begin{eqnarray}
D(z) = G_0 + \sum_{n=1}^{n_{\rm max}} \left[ e^{nz} G_n + e^{-nz} G_n^{*} \right] 
\end{eqnarray}
with $n_{\rm max} = \infty$, where $G_n$ is the sum of terms whose winding number is $n$ in the positive direction, and $G_n^*$ is those of winding number $n$ in the negative direction.
The first and second derivatives of $D(z)$ with respect to $z$ are 
\begin{eqnarray}
\frac{\partial D(z)}{\partial z} = \sum_{n=1}^{n_{\rm max}} \left[n e^{nz} G_n -n e^{-nz} G_n^{*} \right], \hspace{5mm}
\frac{\partial ^2 D(z)}{\partial z^2} &=& \sum_{n=1}^{n_{\rm max}} \left[n^2 e^{nz} G_n +n^2 e^{-nz} G_n^{*} \right] .
\end{eqnarray}
The saddle point $z_0=x_0+iy_0$ is defined as 
\begin{eqnarray}
\frac{\partial D}{\partial z}(z_0) = \sum_{n=1}^{n_{\rm max}} \left[n e^{n z_0} G_n -n e^{-n z_0} G_n^{*} \right] = \rho.
\end{eqnarray}

For the case that $D(z)$ can be approximated by $n_{\rm max}=1$, i.e. 
$D(z) = G_0 + e^{z} G_1 + e^{-z} G_1^{*}$.
The analysis is almost the same as the leading-order analysis.
Because $\rho$ is a real number, 
the complex phase of $(e^{iy_0} G_1 -e^{-iy_0} G_1^{*})$ is zero. 
Then, the imaginary part of the saddle point $y_0$ is determined as
\begin{eqnarray}
\tan y_0 = -\frac{{\rm Im} G_1}{{\rm Re} G_1} 
=- {\rm Arg}\ G_1 .
\end{eqnarray}
Then, the real part $x_0$ is given by 
\begin{eqnarray}
e^{x_0} |G_1| - e^{-x_0} |G_1| = \rho.
\end{eqnarray}
The real part of the saddle point $x_0$ is 
\begin{eqnarray}
x_0 = \mathrm{arcsinh} \left( \frac{\rho}{2|G_1|} \right).
\end{eqnarray}
$D(z_0)$ and $\partial^2 D/ \partial z^2 (z_0)$ at the saddle point are real numbers,  
\begin{eqnarray}
D(z_0)= F_0 + (e^{x_0} + e^{-x_0}) |G_1|, \hspace{5mm}
\frac{\partial^2 D}{\partial z^2} (z_0) =  (e^{x_0} + e^{-x_0}) |G_1|.
\end{eqnarray}
Substituting $(x_0, y_0)$, $D(z_0)$ and $\partial^2 D/ \partial z^2 (z_0)$ into Eqs.~(\ref{eq:f}) and (\ref{eq:theta}).
$-\partial \ln Z_C(T, V \rho)/ \partial \rho $ can be computed as well as the leading-order calculation.

The convergence of the hopping parameter expansion is investigated in Ref.~\cite{Wakabayashi:2021eye}.
In the hopping parameter expansion, the nonzero contribution of $G_n$ appears from $O(\kappa^{n N_t})$ and $G_n$ with $n \ge 2$ is much smaller than $G_1$ in relatively low-order terms of $\kappa$.
Thus, $D(z)$ is well approximated by $n_{\rm max}=1$ for relatively small $\kappa$.
 In such cases, the sign problem does not appear at all.
However, as the quark mass decreases, the contributions from higher-order terms of $\kappa$-expansion increase.
Therefore, it is necessary to consider $G_n$ with $n \ge 2$. 
The complex phase for $n_{\rm max} \ge 2$ is not as easy to remove as when $n_{\rm max} = 1$.
Then, the complex phase of $D(z_0)$ and $\partial^2 D/ \partial z^2 (z_0)$ must be taken into account.

\section{Application to $SU(3)$ lattice gauge theory}
\label{sec:su3ap}

In the discussion of avoiding sign problems in $U(1)$ gauge theory, the $U(1)$ center symmetry is essentially important, while the center symmetry is $Z_3$ for $SU(3)$ gauge theory.
Here, we discuss the probability distribution of the Polyakov loop on the complex plane in $SU(3)$ gauge theory.
To clarify this argument, we should consider the Polyakov loop without spatial averaging (local Polyakov loop) and their spatial average $\Omega$ (averaged Polyakov loop) separately.

\subsection{Distribution of local Polyakov loop}

\begin{figure}[tb]
\begin{center}
\vspace{0mm}
\includegraphics[width=7.3cm]{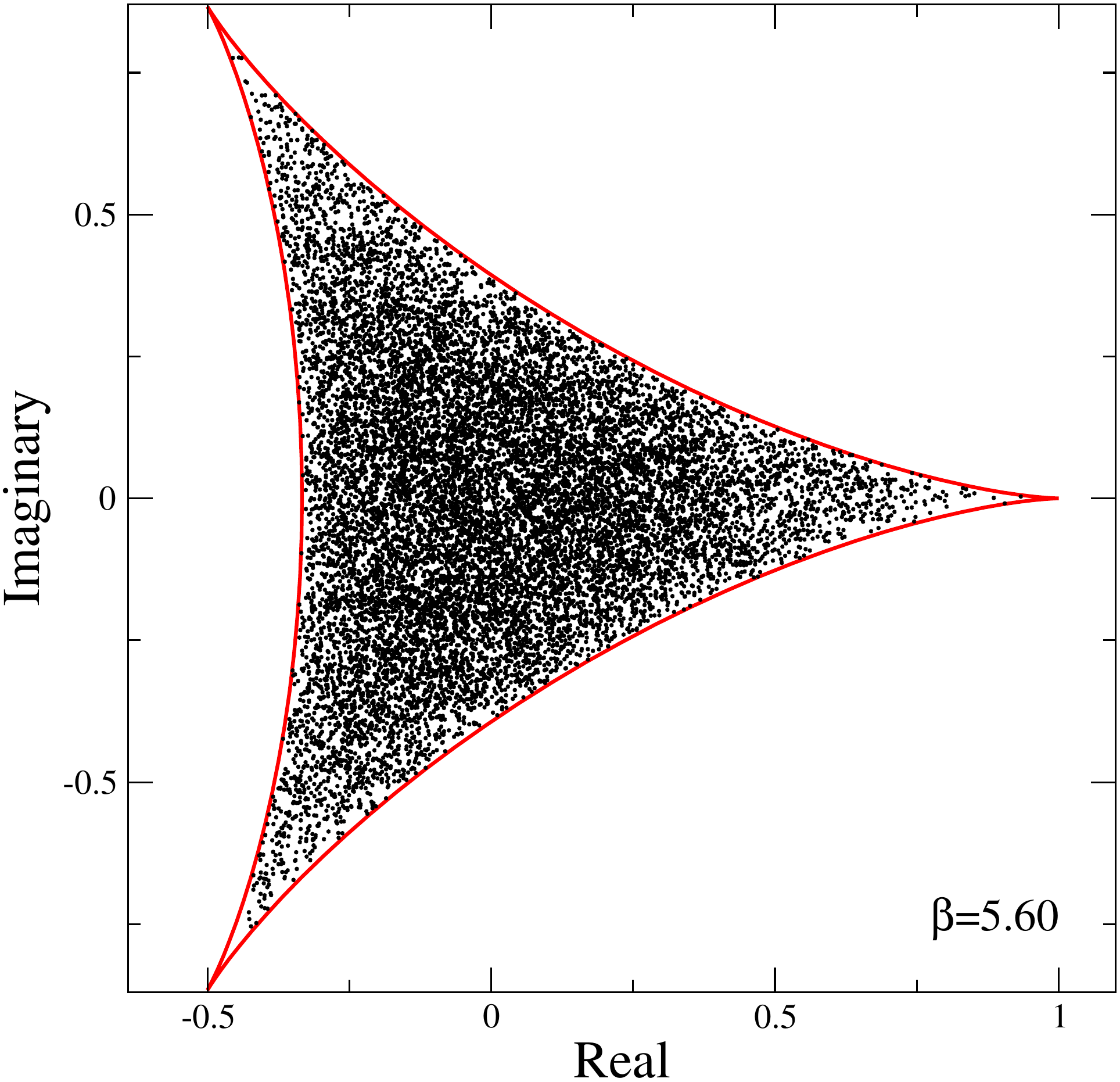}
\hspace{3mm}
\includegraphics[width=7.3cm]{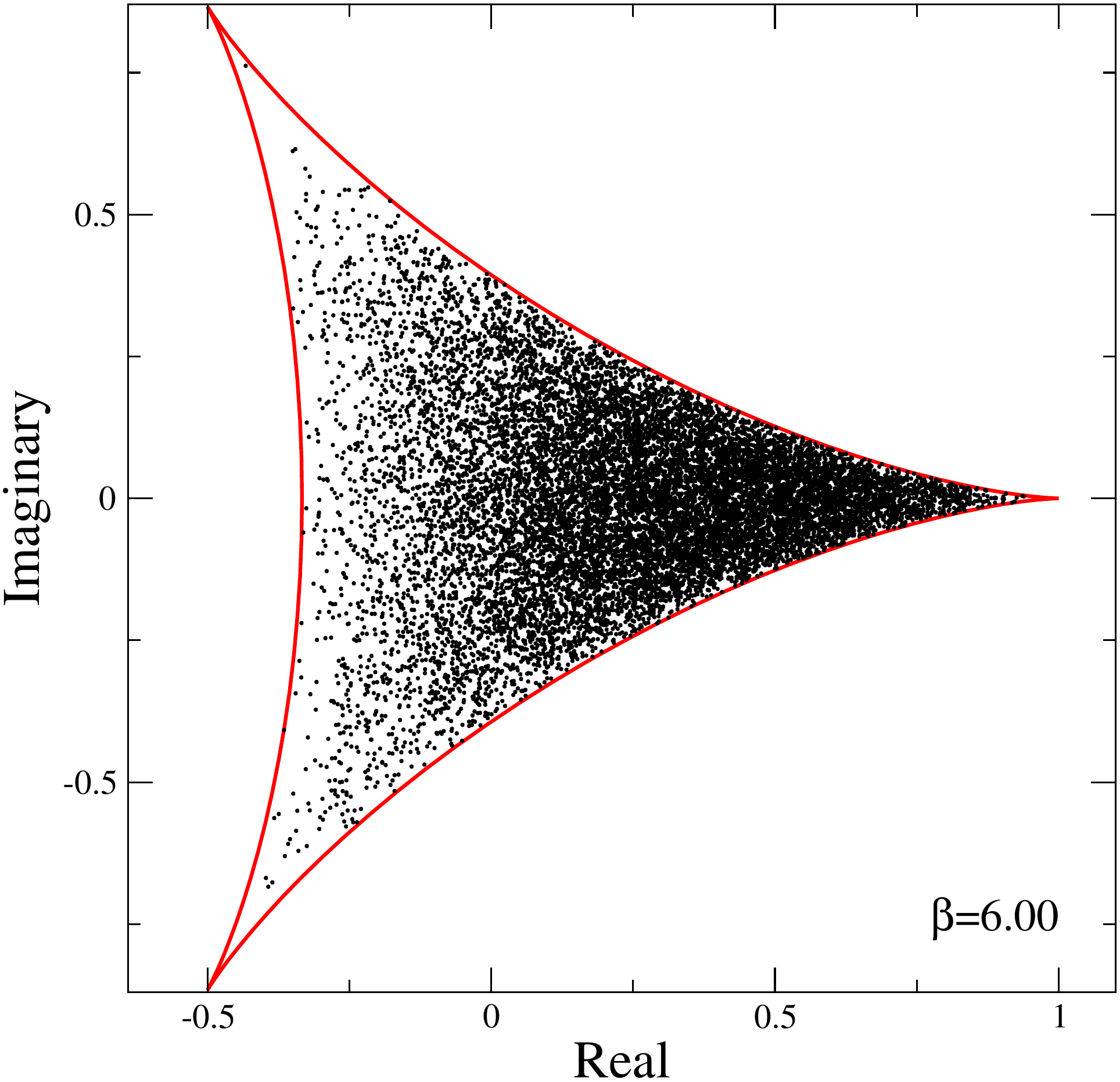}
\vspace{0mm}
\caption{The distribution of the local Polyakov loop in a complex plane on one configuration of $SU(3)$ gauge theory generated at $\beta = 5.60$ (left) and $\beta = 6.00$ (right).}
\label{fig:pl5660}
\end{center}
\end{figure}

First, we discuss the distribution of the Polyakov loop at one spatial point.
We define an element of $SU(3)$ group as 
$U=e^{iH}$. $H$ is a traceless Hermitian matrix, i.e. $\mathrm{tr} H=0$ and $H=H^{\dag}$.
We diagonalize $H$ as $H=V \Lambda V^{\dag}$, 
where $V$ is a unitary matrix and $\Lambda$ is a diagonal matrix with the diagonal elements $(\lambda_1, \lambda_2, \lambda_3)$.
Because $H=H^{\dag}$, $\lambda_1, \lambda_2$ and $\lambda_3$ are real numbers, 
%i.e. $\lambda_i = \lambda_i^*$, 
and $\mathrm{tr} H= \mathrm{tr} \Lambda = \lambda_1 + \lambda_2 + \lambda_3 =0$.
Since the Polyakov loop is the trace of a $SU(3)$ matrix divided by 3, 
$\mathrm{tr} U /3,$ the trace can be rewritten as
\begin{eqnarray}
\mathrm{tr} U &=& \mathrm{tr} \left( e^{iH} \right) 
= \mathrm{tr} \sum_{n=1}^{\infty} \frac{(iH)^n}{n!}
= \sum_{n=1}^{\infty} \frac{i^n}{n!} \mathrm{tr} \left( H^n \right)
= \sum_{n=1}^{\infty} \frac{i^n}{n!} \mathrm{tr} \left( \Lambda^n \right)
\nonumber \\
&=& \sum_{n=1}^{\infty} \frac{i^n}{n!} (\lambda_1^n+\lambda_2^n+\lambda_3^n)
=e^{i \lambda_1} + e^{i \lambda_2} + e^{i \lambda_3} .
\end{eqnarray}
The complex number that can be the value of $\mathrm{tr} U$ is in the triangle drawn by the red lines in Fig.~\ref{fig:pl5660}. 
The square of the absolute value of $\mathrm{tr} U$ is given as
\begin{eqnarray}
|\mathrm{tr} U|^2 &=& (e^{i \lambda_1} + e^{i \lambda_2} + e^{i \lambda_3})
(e^{-i \lambda_1} + e^{-i \lambda_2} + e^{-i \lambda_3})
\nonumber \\
&=& 3+ e^{i (\lambda_1 - \lambda_2)} + e^{i (\lambda_1 - \lambda_3)} + e^{i (\lambda_2 - \lambda_1)} + e^{i (\lambda_2 - \lambda_3)} + e^{i (\lambda_3 - \lambda_1)} + e^{i (\lambda_3 - \lambda_2)}
\nonumber \\
&=& 3+2 \cos (\lambda_1 - \lambda_2) +2 \cos (\lambda_2 - \lambda_3) +2 \cos (\lambda_3 - \lambda_1) .
\end{eqnarray}
The condition for maximizing $|\mathrm{tr} U|^2$ is 
$\lambda_1 = \lambda_2 = \lambda_3$ in $\mathrm{mod} \ 2 \pi$.
Because 
$\lambda_1 + \lambda_2 + \lambda_3 =0$, $\lambda_1 = \lambda_2 = \lambda_3 = 0$, 
$2 \pi/3$ or $4 \pi/3$ in $\mathrm{mod} \ 2 \pi$.
Then, $|\mathrm{tr} U| =3$, and 
the Polyakov loop is $\mathrm{tr} U /3 = (1,0), (-1/2, \sqrt{3}/2)$, or $(-1/2, -\sqrt{3}/2)$.
The condition for the boundary is $\lambda_1 = \lambda_2 \neq \lambda_3$,
$\lambda_2 = \lambda_3 \neq \lambda_1$ or 
$\lambda_3 = \lambda_1 \neq \lambda_2$.
In the case of $\lambda_1 = \lambda_2 \neq \lambda_3$, for example, 
$|\mathrm{tr} U|^2 = 5+4 \cos x$ as a function of $x= \lambda_3 - \lambda_1$.
Then, 
$\mathrm{tr} U /3 = (2 e^{-i x/3} +e^{2i x/3})/3$, 
which is the red curve in Fig.~\ref{fig:pl5660}.

We perform simulations of $SU(3)$ gauge theory (quenched QCD) on a lattice with the size $24^3 \times 4$.
The phase transition point is $\beta_c =5.69138(3)$ on the $24^3 \times 4$ lattice \cite{Saito:2011fs}.
Focusing on one configuration after sufficient thermalization, we investigate the distribution of the local Polyakov loop.
In the left and right panels of Fig.~\ref{fig:pl5660}, we plot the local Polyakov loop at each point on a configuration in the confinement phase $\beta=5.6$ and the deconfinement phase $\beta=6.0$, respectively.
The distributions are in the triangle drown by red curves.
In the confinement phase, the distribution of the local Polyakov loops at each point is $Z_3$ symmetric.
On the other hand, the $Z_3$ symmetry in the distribution of the local Polyakov loop is broken in the deconfinement phase.

\subsection{Distribution of averaged Polyakov loop}

Next, we discuss the distribution of the averaged Polyakov loop.
We map the real and imaginary parts of the local Polyakov loop to two-dimensional plane of $\vec{x}=(x, y)=(r \cos \theta, r \sin \theta)$.
We calculate the variance in a certain direction.
Let $\vec{e}$ be the unit vector in the direction perpendicular to that direction, and let $\eta$ be the angle between $\vec{e}$ and the $x$-axis.
The distance from the straight line drawn in the $\vec{e}$ direction from the origin is 
$| \vec{x} \times \vec{e} | = r \sin (\theta - \eta)$.
The variance is $\int | \vec{x} \times \vec{e} |^2 P(\vec{x}) \ d\vec{x}$, where $P(\vec{x})$ is the probability that the complex value of the local Polyakov loop will be $\vec{x}$.
Here, we assume that the probability distribution is $Z_N$ symmetric.
Then, the generation probabilities of $Z_N$ symmetric points: $\vec{x}_i$ with $i=1, 2, \cdots, N$ satisfying  
$\vec{x}_1 + \vec{x}_2 + \cdots + \vec{x}_N =0$ and $|\vec{x}_1| = |\vec{x}_2| = \cdots =  |\vec{x}_N| =r$, are equal, i.e. 
$P(\vec{x}_1) = P(\vec{x}_2) = \cdots = P(\vec{x}_N)$.
The variance is 
\begin{eqnarray}
\int | \vec{x} \times \vec{e} |^2 P(\vec{x}) \ d\vec{x} 
= \frac{1}{N} \sum_{i=1}^N \int | \vec{x}_i \times \vec{e} |^2 P(\vec{x}_i) \ d\vec{x}_i
= \int \frac{1}{N} \sum_{i=1}^N | \vec{x}_i \times \vec{e} |^2 P(\vec{x}_1) \ d\vec{x}_1.
\end{eqnarray}
We calculate the sum for symmetric points before integrating over $\vec{x}$, 
\begin{eqnarray}
\chi^2 &=& 
\frac{1}{N} \sum_{i=1}^N | \vec{x}_i \times \vec{e} |^2 =
\frac{1}{N} \sum_{i=1}^N r^2 \sin^2 (\theta_i - \eta)
= \frac{r^2}{N} \sum_{i=1}^N \frac{1-\cos 2(\theta_i - \eta)}{2} 
\nonumber \\
&=&  \frac{r^2}{N} \left( \frac{N}{2}-\frac{1}{2}\sum_{i=1}^N \cos 2(\theta_i - \eta) \right)
= r^2 \left( \frac{1}{2}-\frac{1}{2N} \mathrm{Re} \left[ e^{-2i \eta} \sum_{i=1}^N  e^{2i \theta_i} \right] \right), 
\end{eqnarray}
where $\vec{x}_i = (x_i, y_i)=(r \cos \theta_i, r \sin \theta_i)$. 
Because $\theta_n$ is given by $\theta_n = \theta_1 + 2 \pi (n -1)/N$, 
\begin{eqnarray}
\sum_{n=1}^N e^{2i \theta_n} = \sum_{n=1}^N e^{2i (\theta_1 + 2 \pi (n-1)/N)} = e^{2i \theta_1} \sum_{n=1}^N  e^{4i \pi (n-1)/N} =0, 
\end{eqnarray}
except for $N =2$. Hence, $\chi^2 = r^2 /2 = |\vec{x}_1|^2/2$ for $N \geq 3$, and
\begin{eqnarray}
\int | \vec{x} \times \vec{e} |^2 P(\vec{x}) \ d\vec{x} 
= \int \frac{|\vec{x}|^2}{2} P(\vec{x}) \ d\vec{x} .
\end{eqnarray}
The result of the variance does not depend on $\eta$.
This means that the variances in the real axis direction and the imaginary axis direction are the same when the probability distribution is $Z_N$ symmetric.

For the case that the distribution of the local Polyakov loop has $Z_N$ symmetry with $N \geq 3$, the variances of the real and imaginary parts are the same. 
When the volume is large enough, the distributions of the real and imaginary parts of the averaged Polyakov loop are Gaussian by the central limit theorem.
The width of the Gaussian distribution is the same for the real and imaginary parts.
Then, the probability distribution of the averaged Polyakov loop 
$\Omega = |\Omega| e^{i \phi}$ is
\begin{eqnarray}
W(|\Omega|, \phi) \approx C e^{-\alpha (\mathrm{Re} \Omega)^2} e^{-\alpha (\mathrm{Im} \Omega)^2} 
= C e^{-\alpha |\Omega|^2}, 
\end{eqnarray}
where the parameter $\alpha$ is inversely proportional to the variance of the local Polyakov loop distribution and $C$ is the normalization constant.
The distribution does not depend on the complex phase $\phi$ and has $U(1)$ symmetry.

To summarize the above discussion, the distribution of the local Polyakov loops is $Z_3$ symmetric in the confinement phase. 
Then, the probability distribution of the spatially averaged Polyakov loop $\Omega$ is $U(1)$ symmetric in the volume infinity limit, which is not $Z_3$ symmetric.
Since the distribution of $\Omega$ is $U(1)$ symmetric, the method of avoiding the sign problem used in $U(1)$ gauge theory can be applied.
On the other hand, in the deconfinement phase of $SU(3)$ gauge theory, 
$Z_3$ symmetry is spontaneously broken as seen in the right panel of Fig.~\ref{fig:pl5660}.
Then, the probability distribution of the averaged Polyakov loop is not $U(1)$ symmetric.
Thus, the method to avoid the sign problem in $U(1)$ gauge theory cannot be applied.
However, the sign problem is not serious in the deconfinement phase. 
The complex phase distribution can be well approximated by the Gaussian distribution. 
In such a case, the sign problem may be avoided by the method used in Refs.~\cite{Ejiri:2007ga,Ejiri:2009hq,Nakagawa:2011eu,Nakagawa:2012yx,Saito:2013vja}.

\section{Conclusions}
\label{sec:conclusion}

We studied the probability distribution function of particle density. 
The probability distribution function is obtained by calculating the canonical partition function fixing the number of particles from the grand partition function. 
However, if the system has the center symmetry on a finite lattice, 
the canonical partition function is zero when the number of particles is not a multiple of three for $SU(3)$ gauge theory.
For $U(1)$ gauge theory, the canonical partition function is zero except when the particle number is zero.
Then, the probability distribution function is zero for these cases. 
This situation is natural in the confined phase, but is unacceptable in the deconfinement phase because there should be states of various particle numbers.

This problem is essentially the same as the problem that the expectation value of the Polyakov loop is always zero when calculating with finite volume, due to the center symmetry. 
To solve this problem, it is necessary to add an infinitesimal external field to break the symmetry and take the limit of infinite volume.
Moreover, in the case of $U(1)$ gauge theory, the sign problem can be solved using the $U(1)$ center symmetry at the same time.

We performed numerical simulations of $U(1)$ lattice gauge theory near the deconfinement phase transition point.
When the dynamical fermions are heavy, we actually demonstrated that the calculation of the probability distribution function at finite density is possible using the method proposed in this study. 
We calculated the derivative of the canonical partition function using a saddle point approximation \cite{Ejiri:2008xt}, and found that our method to avoid the sign problem works well.
From the canonical partition function, we calculated $\mu /T$ as a function of density $\rho$.
Then, the nature of the phase transition can be investigated.

The application of this method to QCD ($SU(3)$ gauge theory) was discussed.
If the distribution of the local Polyakov loops is $Z_3$ symmetric in the confinement phase, the probability distribution of the spatially averaged Polyakov loop $\Omega$ is $U(1)$ symmetric when the spatial volume is sufficiently large. 
Then, the method of avoiding the sign problem and solving the problem of the center symmetry used in $U(1)$ gauge theory can be applied.
On the other hand, in the deconfinement phase, the probability distribution of the averaged Polyakov loop is not $U(1)$ symmetric.
Thus, the method to avoid the sign problem in $U(1)$ gauge theory cannot be applied.
However, the sign problem is not very serious in the deconfinement phase. 

%\vspace{5mm}
\subsection*{Acknowledgments} 
%\paragraph*{Acknowledgments} 
The author thanks the members of the WHOT-QCD Collaboration for useful discussions.
This work was supported by JSPS KAKENHI Grants No. JP21K03550, No. JP20H01903,
and No. JP19H05146, and the Uchida Energy Science Promotion Foundation.
The author also thanks the Yukawa Institute for Theoretical Physics at Kyoto University for the workshops YITP-W-19-09.

\end{document}